%% file: main.tex
\pdfoutput=1
\documentclass[conference]{IEEEtran}
\IEEEoverridecommandlockouts
\usepackage{cite}
\usepackage{adjustbox}
\usepackage{amssymb,lineno}
\usepackage{algorithm}
\usepackage{algorithmic}
\usepackage{multirow}
\usepackage[tight,footnotesize]{subfigure}
\usepackage{epsfig}
\usepackage{amssymb}
\usepackage{footmisc}
\usepackage{footnpag}
\usepackage{amsmath}
\usepackage{paralist}
\usepackage{color}
\usepackage[dvipsnames]{xcolor}
\usepackage{perpage}
\usepackage{verbatim}
\usepackage{enumitem}
\usepackage{tikz}
\usepackage{listings}
\usepackage{graphicx}
\usepackage{soul}
\usepackage{fancyhdr}
\def\BibTeX{{\rm B\kern-.05em{\sc i\kern-.025em b}\kern-.08em
    T\kern-.1667em\lower.7ex\hbox{E}\kern-.125emX}}
    
\setlist[itemize]{topsep=\parskip}

\begin{document}

\newcommand\xin[1]{{\color{cyan}{Xin: #1}}}
\newcommand\sheng[2]{{\color{red}{Sheng: #1}}}

\pagestyle{plain}

\title{Dynamic Quality Metric Oriented Error-bounded Lossy Compression for Scientific Datasets}

\author{\IEEEauthorblockN{Jinyang Liu,\IEEEauthorrefmark{1}
Sheng Di,\IEEEauthorrefmark{2}
Kai Zhao,\IEEEauthorrefmark{5}
Xin Liang,\IEEEauthorrefmark{3}
Zizhong Chen,\IEEEauthorrefmark{1}
Franck Cappello\IEEEauthorrefmark{2}\IEEEauthorrefmark{4}}
\IEEEauthorblockA{\IEEEauthorrefmark{1}University of California, Riverside, CA, USA}
\IEEEauthorblockA{\IEEEauthorrefmark{2}Argonne National Laboratory, Lemont, IL, USA}

\IEEEauthorblockA{\IEEEauthorrefmark{3}
University of Kentucky, Lexington, KY, USA}
\IEEEauthorblockA{\IEEEauthorrefmark{4}
University of Illinois at Urbana-Champaign, Urbana, IL, USA}
\IEEEauthorblockA{\IEEEauthorrefmark{5}
University of Alabama at Birmingham, Biringham, AL, USA}
jliu447@ucr.edu, sdi1@anl.gov, kzhao@uab.edu, xliang@cs.uky.edu, chen@cs.ucr.edu, cappello@mcs.anl.gov
\thanks{Corresponding author: Sheng Di, Mathematics and Computer Science Division, Argonne National Laboratory, 9700 Cass Avenue, Lemont, IL 60439, USA}
}

\maketitle

\thispagestyle{fancy}
\lhead{}
\rhead{}
\chead{}
\lfoot{\footnotesize{SC22, November 13-18, 2022, Dallas, Texas, USA \newline 978-1-6654-5444-5/22/\$31.00 \copyright 2022 IEEE}} 
\rfoot{}
\cfoot{}
\renewcommand{\headrulewidth}{0pt} \renewcommand{\footrulewidth}{0pt}

\begin{abstract}
With the ever-increasing execution scale of high performance computing (HPC) applications, vast amounts of data are being produced by scientific research every day. Error-bounded lossy compression has been considered a very promising solution to address the big-data issue for scientific applications because it can significantly reduce the data volume with low time cost meanwhile allowing users to control the compression errors with a specified error bound. The existing error-bounded lossy compressors, however, are all developed based on inflexible designs or compression pipelines, which cannot adapt to diverse compression quality requirements/metrics favored by different application users. In this paper, we propose a novel dynamic quality metric oriented error-bounded lossy compression framework, namely \textit{QoZ}. The detailed contribution is three-fold. (1) We design a novel highly-parameterized multi-level interpolation-based data predictor, which can significantly improve the overall compression quality with the same compressed size. (2) We design the error-bounded lossy compression framework QoZ based on the adaptive predictor, which can auto-tune the critical parameters and optimize the compression result according to user-specified quality metrics during online compression. (3) We evaluate QoZ carefully by comparing its compression quality with multiple state-of-the-arts on various real-world scientific application datasets. Experiments show that, compared with the second-best lossy compressor, QoZ can achieve up to 70\% compression ratio improvement under the same error bound, up to 150\% compression ratio improvement under the same PSNR, or up to 270\% compression ratio improvement under the same SSIM.

\end{abstract}


\input{tex/1_introduction.tex}
\input{tex/2_relatedwork.tex}

\input{tex/3_problem.tex}

\input{tex/4_design.tex}

\input{tex/5_predictor.tex}
\input{tex/6_optimization.tex}

\input{tex/7_evaluation.tex}
\input{tex/8_conclusion.tex}

\section*{Acknowledgments}
This research was supported by the Exascale Computing Project (ECP), Project Number: 17-SC-20-SC, a collaborative effort of two DOE organizations – the Office of Science and the National Nuclear Security Administration, responsible for the planning and preparation of a capable exascale ecosystem, including software, applications, hardware, advanced system engineering and early testbed platforms, to support the nation’s exascale computing imperative. The material was supported by the U.S. Department of Energy, Office of Science, Advanced Scientific Computing Research (ASCR), under contract DE-AC02-06CH11357, and supported by the National Science Foundation under Grant OAC-2003709, OAC-2104023 and OAC-2153451. We acknowledge the computing resources provided on Bebop (operated by Laboratory Computing Resource Center at Argonne) and on Theta and JLSE (operated by Argonne Leadership Computing Facility).

\bibliographystyle{IEEEtran}
\bibliography{references}

\end{document}

%% file: tex/1_introduction.tex
\section{Introduction}
\label{sec:introduction}

Modern high-performance computing (HPC) applications across different scientific domains easily produce vast volumes of data for post hoc analysis because of the extremely large scale of the execution required. The Gyrokinetic Toroidal Code (GTC) \cite{GTC} -- an application simulating magnetic particles' movement in confined fusion plasma, for example, may generate many petabytes of data over the course of a few hours \cite{osti_1558473}. Climate applications such as Community Earth System Model (CESM) \cite{cesm,cesm-eval-hpdf14} may also easily generate vast amounts of data every few seconds during one simulation run \cite{gmd-9-4381-2016}. 

In order to mitigate the serious burden of storage or transfer of such a vast amount of scientific data, compression techniques have been studied for years. Lossless compressors are not suitable for compressing scientific data because of their fairly low compression ratios (generally around 2:1 or lower) \cite{sz3}. In contrast, error-bounded lossy compression has been commonly recognized as the most effective data reduction method for scientific application data\cite{mdz, szx, ftsz, tao2018improving}. Not only can it significantly reduce the data volume with a compression ratio of several hundred or even higher, but it can also control the data distortion based on user-specified error bounds.  

The existing error-bounded lossy compressors, however, all have inflexible designs, which cannot adapt to users' diverse requirements for reconstructed data quality. The most popular error control mode (e.g., supported by SZ \cite{sz16,sz17,Xin-bigdata18,sz3modular} and ZFP \cite{zfp}) is absolute error bound: i.e., the difference between original data and reconstructed data must be confined within a constant threshold for each data point. In addition to the error bound constraint, the users may care about the rate-distortion (i.e., the relationship between compression ratio vs. some specific quality metric value) according to their post hoc analysis. Rate distortion may involve different quality metrics in practice. For instance, peak signal-to-noise ratio (PSNR) \cite{z-checker} (equivalent with normalized root mean squared error (NRMSE)) is a common quality metric to assess the overall statistical distortion of the data \cite{isabela,sz17,sz-psnr,Baker-Climate17,use-case,psnr-usecase}. Climate researchers \cite{Baker-Climate17}, for example, often use NRMSE to evaluate the quality of reconstructed data generated by lossy compressors. Structural Similarity Index (SSIM) \cite{ssim} is a perceptual metric that quantifies the visualization quality for a reconstructed data snapshot, which has also been widely used to assess the reconstructed data quality \cite{dssim,Baker-Climate17,evaluate-impact-climate,use-case,z-checker}. Low auto-correlation (AC) of compression errors \cite{z-checker,sz-auto} is often highly preferred by users because it is consistent with the white noise nature. In practice, under the same error bound, the reconstructed data generated by various lossy compression methods often exhibit different levels on these distortion quality metrics, although they have the same maximum compression error. 
Although some lossy compressors (such as MGARD \cite{MGARD} and Fixed-PSNR based compression \cite{sz-psnr}) support preserving different quantity of interest (QoI) metrics (such as L-infinity, L1-norm and L2-norm errors), they are just preserving a threshold of the metric and none of them can dynamically optimize the compression based on diverse quality metrics under a certain error bound. That is, given a particular error bound, the existing lossy compressors always output the compressed and reconstructed data with a fixed compression ratio, which leaves a significant gap for users to control the compression quality on demand. 



In this paper, we propose a novel quality metric oriented error-bounded lossy compression framework, called \textit{QoZ}, which faces several challenging issues, regardless of the compression models used in the study. (1) Combining the user-specified quality metric with error-bounded lossy compression requires an in-depth investigation of various lossy compression models. (2) Based on a specified quality metric, determining which steps or what parameters in the compression are tunable and critical to the overall compression quality is non-trivial. (3) How to optimize the rate distortion with respect to the user-specified quality metric is non-trivial, since in this case, the compression result regards a co-optimization of the compression ratio and the quality metric instead of just maximizing the compression ratio. A straightforward method is using trial-and-error search to run the compressor multiple times with different tunable parameters, which inevitably introduces very expensive computation cost  \cite{underwood2020fraz,underwood2022optzconfig}.

%

To the best of our knowledge, our developed QoZ compression framework is a fresh attempt to adaptively adjust compression quality based on different quality metrics online under a particular error bound. The key contributions are summarized as follows.



\begin{itemize}
    \item We carefully explore and design the best-fit data predictor for building our error-bounded lossy compression framework QoZ. 
    \item We develop an efficient error-bounded lossy compression framework that can dynamically optimize different inclined quality metrics in online compression. To this end, we leverage multiple advanced techniques, including block-wise anchor point structure, multi-level interpolation-based data prediction, level-wise predictor selection, and error bound auto-tuning. 
    \item We evaluate our proposed solution QoZ by using multiple real-world scientific application datasets. Experiments show that QoZ achieves considerable improvements upon the second-best lossy compressor in terms of various quality metrics: e.g., up to 70\% compression ratio improvement under the same error bound, up to $\sim$150\% compression ratio improvement under the same PSNR, or up to $\sim$270\% compression ratio improvement under the same SSIM.  
\end{itemize}

The remaining of the paper is organized as follows. Section \ref{sec:related} discusses the related state-of-the-art work. Section \ref{sec:problemform} formulates the research problem. In Section \ref{sec:design}, we present the overview of the design. In Section \ref{sec:predictor}, we detail the critical part of our design -- data predictor. In Section \ref{sec:optimization}, we describe our predictor optimization strategies in detail. In Section \ref{sec:evaluation}, we present and analyze the evaluation results. Finally, we present the concluding remark and future work in Section \ref{sec:conclusion}. 

%% file: tex/2_relatedwork.tex
\section{Related Work}
\label{sec:related}

In this section, we discuss the related work, which includes two topics: (1) the state-of-the-art error-bounded lossy compressor, and (2) the analysis algorithm to select the best compressor or predictor based on a specific dataset.

There have been many state-of-the-art lossy compressors developed for scientific applications. Basically, they can be split into four categories: prediction-based model, transform-based model, dimension-reduction model and neural network based model. The prediction-based compression model predicts each data point based on its neighbor values or saved coefficients (e.g., using Lorenzo predictor \cite{lorenzo} and linear-regression predictor \cite{Xin-bigdata18}) then use a quantization or similar methods to control the errors within user-specified error bound. Typical examples include FPZIP \cite{lindstrom2017fpzip}, SZ2 \cite{Xin-bigdata18} and MGARD \cite{MGARD}. Recently proposed SZ3 \cite{sz3} leverages dynamic spline interpolation for data prediction, and greatly outperforms other existing lossy compressors in terms of rate-distortion in multiple cases.
The transform-based compression model transforms the data into another coefficient domain such that the coefficient data are much easier to compress because of its fairly high sparsity in nature. One typical example is ZFP \cite{zfp}, which adopts exponent alignment + orthogonal transform + embedded encoding on non-overlapped split small blocks. The dimension-reduction model aims to reduce the dimensions of the dataset by leveraging (high-order) singular vector decomposition (SVD). One typical example is TTHRESH \cite{ballester2019tthresh}. The neural network based compression model \cite{ae-sz,glaws2020deep,liu2021high,hayne2021using} leverages neural networks such as autoencoder \cite{ae} and its variations (VAE \cite{vae}, SWAE \cite{swae} et al.) for data compression and reconstruction. Typical examples are AE-SZ \cite{ae-sz} and the work of Liu et al. \cite{liu2021high}. All the existing error-bounded lossy compressors, however, always have fixed static designs so that none of them allows to optimize the compression quality in terms of specific quality metric adaptively.

Since different lossy compression algorithms exhibit distinct pros and cons on different datasets, several existing works have studied how to select the best compression model, best predictor or parameters at runtime. Lu et al. \cite{lu2018understanding} proposed a compression-selection method by estimating the compression ratio for SZ and ZFP based on a particular relative error bound. Tao et al. \cite{tao2019optimizing} proposed an analysis method that can select the better choice between SZ and ZFP based on the PSNR, which is a very popular metric used in the visualization community. Liang et al. \cite{liang2019significantly} developed a method integrating ZFP compressor as a predictor in SZ compression model and proposed a method to select the best-fit predictor for the whole input data at runtime. 
Zhao et al. \cite{sz-auto} proposed a sampling method to search for the best-fit parameter settings and predictors under the SZ compression model, which can significantly improve the compression quality over SZ2 \cite{Xin-bigdata18} in turn. However, each of them supports only one specific quality metric (such as PSNR and error bound) because of their static analysis, which cannot adapt to diverse quality metrics. Our proposed QoZ overcomes this limitation, having the cabability of online auto-tuning parameters for different user-given quality metrics. Moreover, all these existing methods select the better compressor or predictor for the entire dataset, while our method is able to fine-tune the best-fit solution/predictor for different data points (based on levels) in the dataset, leading to a substantially higher flexibility and compression quality.      

%% file: tex/3_problem.tex
\section{Problem Formulation}
\label{sec:problemform}

In this section, we formulate the research problem to clarify our research objective. Basically, we focus on a dual-objective lossy compression problem: meeting the necessary condition (error bound constraint) meanwhile optimizing the compression result in terms of the user-specified quality metrics. For example, \textit{Rate distortion} is a very common method to assess the lossy compression quality. \textit{Rate} here refers to the \textit{bit-rate}, which is defined as the average number of the bits used to represent a data point after compression. Obviously, the lower the bit rate, the better the compression result. \textit{Distortion} measures the difference between the original data and the decompressed data, and in literature, it is mainly referred to as peak signal-to-noise ratio (PSNR) \cite{z-checker} (to be detailed later). In our work, we extend the concept of the rate-distortion to fit more generic distortion metrics such as SSIM \cite{ssim} and autocorrelation (AC) \cite{z-checker} of compression errors, which is a critical advancement to optimize compression quality based on user's requirement on data fidelity in practice. In the following text, we first briefly introduce the fundamental concept of error-bounded lossy compression then formulate the quality metric oriented compression problem. 

The error-bounded lossy compression is formulated as follows. Given an input data array (denoted by $X$) and a user-specified absolute error bound $e$, the error-bounded lossy compression consists of a compressor $C$ and a decompressor $D$. It generates the compressed data (denoted $Z$) and the decompressed data (denoted $X^{'}$), which strictly respects the error bound (denoted $e$) on each data point. For each data value $d_i$, $|d_i-d_i^{'}| \leq e$ must be satisfied, where $d_i$$\in$$X$ and $d_i'$$\in$$X'$ represent the original data value and decompressed data value, respectively. 

In this paper, we aim to develop a highly parameterized error-bounded lossy compression framework, which can auto-tune the parameters to obtain the best rate-distortion in terms of different quality metrics such as PSNR, SSIM, and AC. We denote the error bound by $e$ and the user-specified quality metric by $T$. For a specific parameter set $\theta$ and parameterized compressor $C_\theta$ and decompressor $D_\theta$, our QoZ can automatically determine $\theta$ according to the optimization problem formulated as follows:

\begin{gather*}
\theta=\underset{\theta}{\arg \textsc{OPT}} \hspace{1mm}T(X,X^{'},Z)\\
s.t. \hspace{2mm} Z=C_\theta(X)\\
\hspace{7mm} X^{'}=D_\theta(Z)\\
\hspace{20mm} |x_i-x_i^{'}|\leq e, \forall x_i \in X 
\end{gather*}
where OPT refers to a optimization operation (e.g., $\max$, $\min$) according to the specific quality metric. 



We describe a few well-known quality metrics as follows (but not limited in practice):
\subsubsection{\textbf{PSNR}} PSNR is a metric commonly used in the rate-distortion evaluation \cite{z-checker}. PSNR measures the data distortion (i.e., compression errors) according to the following formula ($vrange$ means the value range = $\max$(X)-$\min$(X)): 
\def\formulaPSNR{
P\hspace{-0.3mm}S\hspace{-0.3mm}N\hspace{-0.3mm}R =
20\log_{10}\frac{vrange(X)}{\sqrt{mse(X,X')}}
}
\begin{equation}
\label{eq:psnr}
  \formulaPSNR
\end{equation}
\noindent


\subsubsection{\textbf{SSIM}} SSIM \cite{ssim} is a significant metric commonly used to measure the visual quality of decompressed data. The formula for calculating SSIM with input data $X$ and decompressed data $X^{'}$ is: 
\begin{equation}
\label{eq:ssim}
\begin{array}{l}
 SSIM = \frac{1}{N}\sum\nolimits_{i = 1}^N {SSIM_i (X,X')} 
 \end{array}
\end{equation}
$SSIM_i(X,X')$ is the calculation of SSIM for a local sliding window $i$, which is calculated as follows:
\begin{equation}
\label{eq:ssim2}
\begin{array}{l}
SSIM_i(X,X')=\frac{(2\mu_X\mu_{X'}+c_1)(2\sigma_{XX'}+c_2)}{(\mu_X^2+\mu_{X'}^2+c_1)(\sigma_X^2+\sigma_{X'}^2+c_2)}
 \end{array}
\end{equation}
where $\mu$ is the mean, $\sigma$ is the standard variance/covariance. For details, We refer readers to read Wang et al.'s papers \cite{ssim}.
 
\subsubsection{\textbf{Auto-correlation of compression error (AC)}} 
Au-correlation of compression error (AC) is a very important metric concerned by many application users. It is defined as follows. 
\begin{equation}
\label{eq:ac}
\begin{array}{l}
AC = \frac{{E(e_i  - \mu _i )(e_{i + k}  - \mu _{i + k} )}}{{\sigma ^2 }}
 \end{array}
\end{equation}
where $e_i$ denotes the compression error at data point $i$ and $k$ is the lag (or offset) used to calculate the auto-correlation. The lower the AC value, the higher the randomness of the error correlation at adjacent data points. In general, the users expect to have a random error correlation between adjacent data points (i.e., low AC values). 

\begin{figure}[ht]
  \centering
  \raisebox{-1mm}{\includegraphics[scale=0.5]{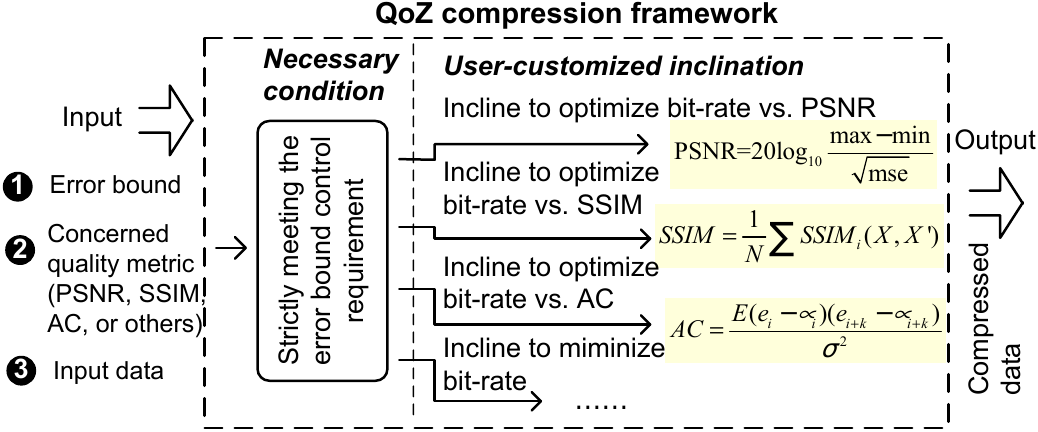}}
  \vspace{-3mm}
  \caption{Error-bounded Quality Metric Oriented Compression Framework}
  \label{fig:QoZFramework}
\end{figure}

We demonstrate our proposed error-bounded quality metric oriented compression framework in Figure \ref{fig:QoZFramework}, by using the above-mentioned quality metrics as an example. 

%% file: tex/4_design.tex
\section{Design Overview}
\label{sec:design}

In this section, we describe the overall design of our quality metric oriented error bounded compression framework -- QoZ. 

QoZ leverages and extends the SZ lossy compression framework, which is the fundamental of multiple existing lossy compressors \cite{sz16, sz17,Xin-bigdata18, sz3}. 
Compared with the existing lossy compressors, the key advantage of QoZ is two-fold. On one hand, not only does QoZ support error-bounding constraints, but at the same time, it allows users to tune the compression results based on a preferred quality metric important for their post hoc analysis. On the other hand, QoZ significantly improves the compression quality through a series of optimization strategies such as level-wise predictor and parameter auto-tuning.   

\begin{figure}[ht]
  \centering
  \raisebox{-1mm}{\includegraphics[scale=0.48]{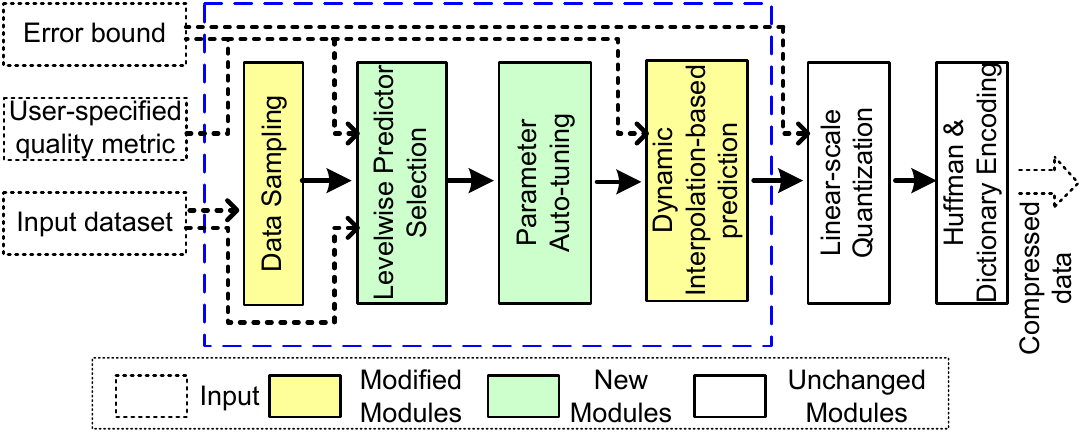}}
  \vspace{-2mm}
  \caption{Design Architecture of QoZ (Quality Oriented Compression)}
  \label{fig:QoZDesign}
\end{figure}

The overall QoZ framework is presented in Figure \ref{fig:QoZDesign}, in which our key design of QoZ is in the blue-dotted rectangle. This part corresponds to the data prediction stage in SZ compression framework, which is the core of the whole compression pipeline. Compared with the latest version of SZ -- SZ3 \cite{sz3}, we developed two new modules (shown as green boxes in the figure), which not only significantly improve the compression quality but also support the optimization of diverse quality metrics. Specifically, our careful observation with masses of real-world scientific datasets indicates that the error controls on high levels of the multi-level interpolation-based predictor is critical to the overall prediction accuracy which directly affects the compression ratio. 
As such, we adopt lossless compression and high precision lossy compression on high interpolation levels to increase the prediction accuracy.
We also integrate the quality metric assessment into the prediction stage to tune the compression quality. That is, QoZ dynamically parameterizes the selected best-fit predictor on different levels according to the user-specified quality metric for optimizing the overall compression quality. 


%% file: tex/5_predictor.tex
\section{Level-adapted Interpolation-based Prediction}
\label{sec:predictor}

In this section, we present our developed level-adapted interpolation-based data predictor used in the QoZ framework. To optimize the compression with various compression quality requirements in terms of different quality metrics, the compression framework needs to be flexible enough to provide multiple compression results under the same error bound constraint or the same compression ratio. As such, a flexible prediction method is critical to the QoZ framework. 

In our design, our prediction stage is based on the spline-interpolation-based data predictor \cite{sz3}, because of the following two reasons:
\begin{itemize}
    \item According to Zhao et al.'s study \cite{sz3}, the spline-interpolation-based predictor can obtain outstanding compression qualities over many other existing lossy compressors such as ZFP and SZ in most cases.
    \item The spline-interpolation-based prediction is executed based on a level-wise architecture, which provides great potential for parameterization and auto-tuning.
\end{itemize}

In what follows, we first introduce the basic design of the spline-interpolation-based data predictor, then describe our developed level-adapted interpolation-based data predictor. 
These newly proposed designs are not only for the target of quality-metric-driven compression but also improve the data prediction accuracy in lossy compression, which will be detailed in the next section. 

\subsection{Basic spline interpolation based predictor}
\label{sec:interp}

\begin{figure}[ht]
  \centering
  \raisebox{-1mm}{\includegraphics[scale=0.94]{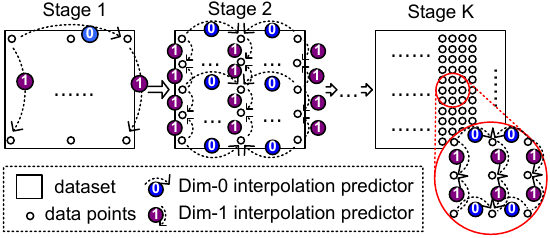}}
  \vspace{-3mm}
  \caption{Illustration of Basic Interpolation Predictor}
  \label{fig:basic-interpolation}
\end{figure}

Compared to the traditional extrapolation methods such as the Lorenzo predictor and regression models such as Linear regression, the interpolation-based predictor can improve the prediction accuracy prominently, especially for smooth datasets, as presented in Zhao et al.'s recent studies \cite{sz3}. In the interpolation-based predictor, the data points in a data array are predicted based on a fixed interpolation method with varied strides, following a fixed propagation policy (as demonstrated in Figure \ref{fig:basic-interpolation} based on a 2D example). The entire prediction procedure starts with the first data point (see Stage 1 in the figure), which will be used to predict large-stride data points through the whole data array, followed by a linear-scale quantization to make sure the reconstructed value is close to the true data value within the expected error bound. Then, more data points would be predicted and quantized along another dimension alternatively, as demonstrated in the figure, until all the data points are covered (see Stage K in the figure). Note that each interpolation operation has to use the reconstructed data values (i.e., the approximated values after prediction+quantization on that data point) instead of the original data values, in order to make sure that the reconstructed data during the decompression would definitely respect the expected error bound. 

\subsection{Level-adapted spline interpolation-based prediction in QoZ}

Our developed interpolation-based data predictor in QoZ eliminates several critical limitations of the basic interpolation-based predictor.

\subsubsection{Improving prediction accuracy by avoiding long-range interpolation}

\begin{figure}[ht]
  \centering
  \raisebox{-2mm}{\includegraphics[scale=0.43]{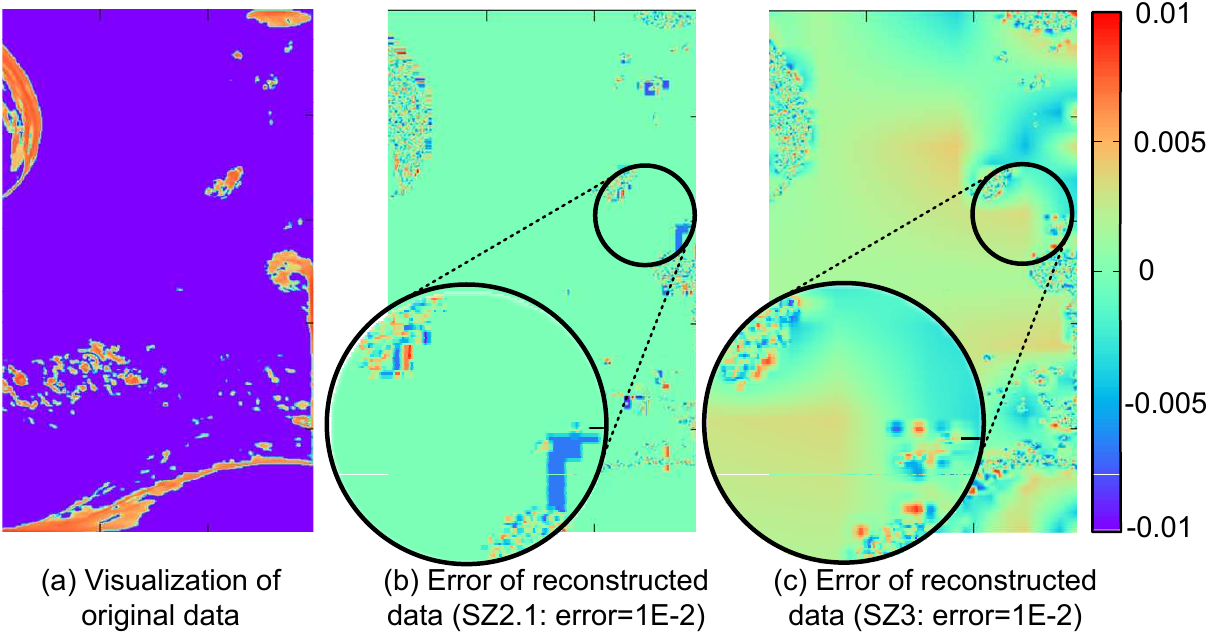}}
  \vspace{-3mm}
  \caption{Visualization of original data and compression error (Hurricane Cloud)}
  \label{fig:cloudvis}
\end{figure}

The first serious issue in the basic interpolation-based predictor is that it suffers from considerably low accuracy in long-range interpolation. 
As mentioned previously, the basic interpolation-based prediction method is executed in the order from large strides to small strides. Since it does not control the maximum stride length, the prediction accuracy would be fairly low when the interpolation spans a long distance in the data array. This situation turns even worse especially when the data exhibits different smoothness or patterns in different areas. In Figure \ref{fig:cloudvis}, we use an example to illustrate this serious situation of the SZ3 which adopts the basic interpolation-based prediction method. We can clearly observe more artifacts in the compression errors generated by SZ3 \cite{sz3} than by SZ2.1 \cite{Xin-bigdata18} (using block-wise linear regression and Lorenzo predictor for data prediction) under the same absolute error bound of 1E-2. This is mainly due to the fact that the interpolation method cannot predict the distant data values accurately in SZ3, while SZ2 always predict data points with their neighbors. 

QoZ leverages grid-wise anchor points to avoid those inaccurate long-range interpolations, which mitigates the inaccurate prediction issue effectively. Specifically, for interpolation, anchor points are data points that are considered to be known in advance and losslessly encoded and saved. 
These anchor points split the whole data array into many blocks and all other data points would be predicted/reconstructed by other points within a certain range, using the multi-level based interpolation method. We note that the storage overhead introduced by saving the losslessly compressed anchor points would be nearly negligible if we set an appropriate stride for the anchor point grid. The key advantage of utilizing anchor points is that it may greatly improve the quality of the distortion metric, which will be presented later on in Section \ref{sec:evaluation}.  



\subsubsection{Level-adapted interpolation and error bound auto-tuning}
To have better prediction accuracy, the QoZ data predictor selects the best interpolation method at corresponding levels during the compression. As described previously, the basic interpolation-based prediction method can be decomposed into multiple stages (as shown in Figure \ref{fig:basic-interpolation}). In QoZ design, we treat these stages as non-overlapping levels: stage 1 corresponds to level K, $\cdots$, and stage K corresponds to level 1. We have two important takeaways regarding these different interpolation levels. 
\begin{itemize}
    \item These interpolation levels may have different data patterns or characteristics from each other, which motivates us to adopt different predictors on different levels.
    \item The compression quality of points at higher levels may affect the compression quality of points at lower levels significantly, in that the prediction of points always relies on the decompressed data points at higher levels.
\end{itemize}

Based on the above two critical takeaways, we develop the level-adapted interpolation-based predictor as follows. 

First, the QoZ interpolation-based predictor adopts diverse interpolation methods at different levels. Specifically, the interpolation type includes both linear interpolation and cubic spline interpolation. As mentioned in Section \ref{sec:interp}, the multi-dimensional interpolation method is actually composed of multiple 1D interpolation operations. In a high-dimensional data array (such as 3D), even for the same interpolation type, each interpolation level may also involve multiple dimensions. As such, different permutations of the dimensions may lead to different prediction qualities. As an example, for 3D data, there are 6 different permutations based on the three dimensions (dim0, dim1, and dim2): 012, 021, 102, 120, 201, 210. Accordingly, considering the two types of interpolation, there are a total of 12 prediction methods to select at each level. 

Second, the QoZ interpolation-based predictor sets different error bounds for different levels. Such a design is motivated by the following important observation. In the whole interpolation-based prediction, a large majority of the data points (75\% in 2D case or 87.5\% in 3D case) are at the lowest level (level 1), but they are mostly predicted by the reconstructed data points from higher levels: a total of 25\% of the data points in 2D case or 12.5\% of the data points in 3D case. Therefore, setting a smaller error bound at a higher level may preserve a very good overall prediction accuracy, which improves the compression quality in turn. 
Another important motivation is that having flexible and online-tuned level-wise error bounds makes the metric-driven optimization of lossy compression possible, with which the compressor can dynamically set error bounds to provide different compression results according to different optimization targets.

\begin{figure}[ht]
  \centering
  \raisebox{-1mm}{\includegraphics[scale=0.5]{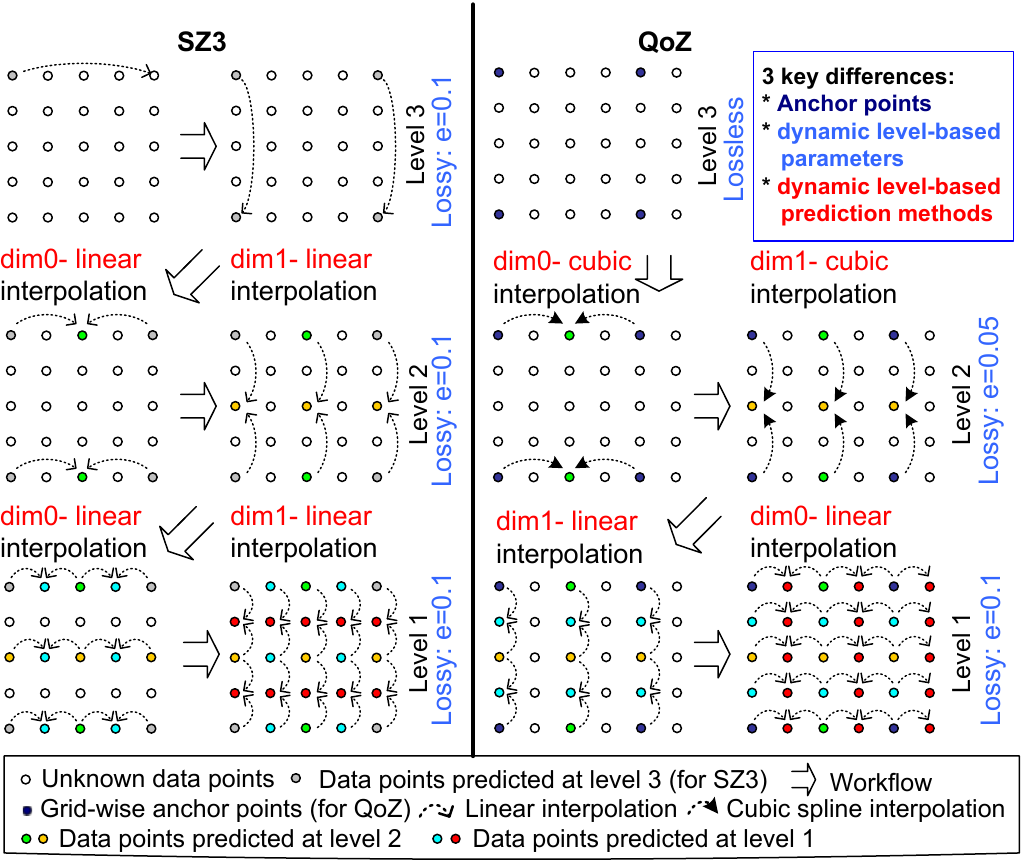}}
  \vspace{-2mm}
  \caption{Illustrating the key difference between QoZ vs. SZ3 (using 2D dataset as an example)}
  \label{fig:levelwise}
\end{figure}

In Figure \ref{fig:levelwise}, we demonstrate the key differences between QoZ and SZ3 using an example (based on a 2D data array). As shown in the figure, there are three key differences: (1) QoZ adopts anchor points that can minimize the error propagation in the interpolation methods from the top level to the lower levels; (2) QoZ dynamically tunes the parameters (i.e., error bounds) at different levels, which can improve compression ratio in turn; (3) QoZ uses a level-adapted interpolation method (highlighted in red font), which can further improve compression ratio in turn. In the example illustrated in the figure, under error bound 0.05 QoZ interpolates along dim 0 $\rightarrow$ dim 1 at level 2 with cubic spline interpolation, while its interpolation dynamically switches to dim 1 $\rightarrow$ dim 0 with linear spline interpolation under error bound 0.1 at level 1. How to select the best-fit prediction method will be discussed later on in Section \ref{sec:interpsec} in detail. 

In our implementation, we design two critical parameters ( $\alpha$ and $\beta$) to tune the level-wise error bounds for the interpolation-based predictors. Specifically, given a global error bound $e$, the error bound for the interpolation level $l$ is:

\begin{equation}
\label{eq:ebl}
e_l=\frac{e}{min(\alpha^{l-1},\beta)} \ (\alpha \geq 1 \ and \ \beta \geq 1)
\end{equation}

The availability and effectiveness of $e_l$ is determined by the following policy:

\begin{itemize}
    \item $e_l \leq e$, $\forall l$. This is to make sure compression errors for all data points must be within the user-set error bound $e$.
    \item $e_1 = e$. This means that the compression of the data points at level 1 which involves 75\% (87.5\%) of data points in the 2D (3D) input uses the maximum acceptable error bound ($e$) to optimize the compression ratio.
    \item $e_{l_1} \geq e_{l_2}$ when $l_1 < l_2$. Since every interpolation has to use lossy reconstructed data instead of the original data (to respect error bound strictly during decompression), the data reconstruction errors would be propagated to all data points at lower levels. Thus, the error bound at higher levels should be more accurate than that at lower levels. 
\end{itemize}
 It is worth noting that since QoZ is a dynamic quality-metric-driven lossy compressor, there would be multiple choices for $\alpha$ and $\beta$ based on the same input dataset and user-set error bound, because of various user-specified quality metrics to target.
 In Section \ref{sec:paramtuning}, we will present how values of $\alpha$ and $\beta$ are determined during the online compression.

%% file: tex/6_optimization.tex
\section{Online Optimization of Compression based on User-specified Inclined Quality Metric}
\label{sec:optimization}

In this section we describe our tuning and optimization strategies for the proposed data predictor in section \ref{sec:predictor}. In regard to the above-mentioned parameterized level-wise interpolation-based predictor, there are several great challenging issues to resolve. For example, what type of interpolation operation should be used on each interpolation level? How to set prediction parameters ($\alpha$ and $\beta$ for computing level-wise error bounds) in order to optimize the user-specified quality metric? 


In QoZ, we propose an efficient online tuning method, which not only can select the best-fit predictor and optimize the parameters at different levels based on diverse quality metrics but also has very low execution overhead such that the overall compression performance can still be maintained well. We detail our optimization strategies in the following text. 


    

\subsection{Efficient uniform sampling in QoZ}
\label{sec:sampling}

\begin{figure}[ht]
  \centering
  \raisebox{-1mm}{\includegraphics[scale=0.46]{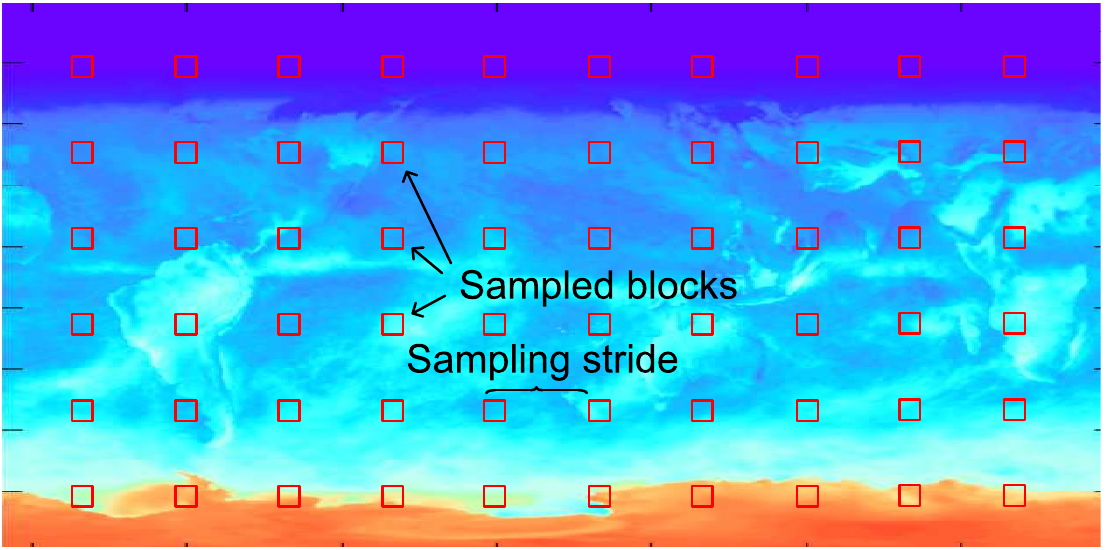}}
  \vspace{-1mm}
  \caption{Illustrating Data Sampling using CESM-ATM dataset (field FSUTOA)}
  \label{fig:sampling}
\end{figure}

In order to control the online analysis overhead, we adopt a data sampling method in QoZ, which plays an important role in reaching a good trade-off between the accuracy of the predictor/parameter selection and the computation overhead of this selection. To this end, the sampled data should be small enough to keep a very low time cost for the analysis and they should be good representatives for the whole input data. As such, we adopt a uniform block-based sampling method, which can catch not only the pattern of the local area in the data but also the global picture of the data effectively. Our sampling method is based on fixed block size and fixed sampling stride, as illustrated in Figure \ref{fig:sampling} using an example based on the CESM-ATM climate simulation dataset. The sampling rate (defined as the percentage of the number of sampled data points over the total number of data points) is determined by both block size and sampling stride. For instance, for a 2D dataset, if the block size is 4$\times$4 and the sampling stride is 10, the sampling rate will be $\frac{4\times4}{10\times10}$=16\%. 
In our parameter tuning over the sampled data, the prediction step is performed separately on each data block while the Huffman and dictionary encoding \cite{zstd} are applied on the entire aggregated quantization bins for accurate bit rate estimation. 

\subsection{Level-adapted selection of best-fit predictor}
\label{sec:interpsec}

Unlike SZ3 \cite{sz3} which uses a fixed interpolation method at different levels throughout the whole data array, QoZ selects and applies the online-determined best-fit interpolation method on different levels with very limited computational overhead compared with the entire compression time. 

We present the pseudo-code of the online selection algorithm in Algorithm \ref{alg:lspsec}. First, QoZ samples data blocks with the method introduced in section \ref{sec:sampling} (line 1). Next, on each interpolation level (line 4), QoZ runs a few trial compression runs with different interpolation/prediction methods (a.k.a., interpolators) over the sampled data blocks. 
As mentioned previously, the candidate interpolators involve two different types of interpolation and different dimension orders for a multi-dimensional input. As the total number of dimension orders grows very fast with the dimension number of the data (e.g. 6 for 3D), being consistent with SZ3\cite{sz3}, QoZ only tests 2 dimension orders: dimension index increasing or decreasing, (e.g. 012 and 210 for 3D) as they cover the best choices in almost all cases.   
Then, the algorithm compares the mean absolute prediction errors ($L_1$) and selects the one with the lowest mean absolute prediction error as the best-fit interpolator for the corresponding level (see line 10$\sim$19). The reason we use absolute error is that it is most closely related to the compression ratio of quantization bins, which was verified in \cite{Xin-bigdata18}. The selection of interpolators does not need to be specifically tuned according to different quality metrics because its purpose is to minimize prediction errors, which both benefit bit rate and quality metrics. Since the interpolator selection is based on the sampled blocks, a tricky situation is that when the sampled block size is smaller than the anchor point stride, the blocks cannot cover some high interpolation levels. To solve this issue, we use the best-fit interpolator selected on the highest level of the sampled blocks ($L$ in algorithm \ref{alg:lspsec}) for all higher levels in the whole data.


\begin{algorithm}
\footnotesize
\caption{Level-adapted selection of best-fit predictor}
\label{alg:lspsec}
\renewcommand{\algorithmiccomment}[1]{/*#1*/}
\begin{flushleft}
\textbf{Input:} Input data $D$, interpolator candidates $I$, sample block size $b$, anchor point stride $s$,sample rate $r$, error-bound $e$.\\ 
\textbf{Output:} Selected interpolator list $I_s$.
\end{flushleft}

\begin{algorithmic}[1]
\STATE $L \leftarrow \log_2\min(b,s)$ \COMMENT{Determine max interpolation level in selection.}
\STATE Sample data blocks $B$ $\leftarrow$ from $D$ with block size $b$ and sampling rate $r$.
\STATE $I_s \leftarrow []$ \COMMENT{Initialized empty result}
\FOR{level $l$ from $L$ to $1$}

\STATE $best\_error \leftarrow +\infty$

\FOR{interpolator $i$ in $I$}
\STATE Based on error bound $e$, run trial compression with $i$ on the level $l$ of $B$. Compute mean $L_1$ prediction error $cur\_error$.
\IF {$cur\_error < best\_error$} 
\STATE  $best\_error \leftarrow cur\_error$ 
\STATE $i_b \leftarrow i$ \COMMENT{Determine the best interpolator on current level.}
\ENDIF
\ENDFOR
\STATE $I_s[l] \leftarrow i_b$ \COMMENT{Set the interpolator on level $l$ as $i_b$}

\ENDFOR
\RETURN $I_s$
\end{algorithmic}
\end{algorithm}

\subsection{Quality metric oriented parameter auto-tuning}
\label{sec:paramtuning}

In this subsection, we describe more technical details about the user-specified quality metric oriented parameter auto-tuning algorithm in QoZ, which is critical to the user-specified quality-metric-driven lossy compression. It includes constructing parameter candidates, online compression result evaluation, and online parameter auto-tuning.

\subsubsection{Constructing parameter candidates}
\label{sec:constructing}

As mentioned previously, we formulate the parameter optimization problem as Formula (\ref{eq:ebl}), which includes two critical parameters $\alpha$ and $\beta$ to determine the best error bound setting for level $l$. According to masses of our experiments with different datasets across various domains, we narrow the best parameter candidates as follows: $\alpha$ =\{1, 1.25, 1.5, 1.75, or 2\} and $\beta$ = \{1.5, 2, 3, or 4\}, because these values cover the optimal or near-optimal settings of $\alpha$ and $\beta$ in most cases without too many pairs to test with. In our algorithm (to be shown later), the optimal combination of the $\alpha$ and $\beta$ will be determined online based on a lightweight compression result evaluation, which brings little computational overhead.  

\subsubsection{Online compression quality evaluation}

Online optimization of the rate-distortion based on a user-specified quality metric with different parameter settings is non-trivial. The key reason is that rate distortion refers to the relationship between bit-rate and the user-specified quality metric. Thus, accurately identifying the rate-distortion for a specific solution generally needs to collect quite a few compression results based on different compression ratios and various levels of data distortions. Specifically, given two different parameter sets (or solutions) each corresponding to a particular compression result, determining which one is better requires a meticulous analysis as described below. Suppose we are checking two solutions: setting I ($\alpha$=1 and $\beta$=1.5) and setting II ($\alpha$=2 and $\beta$=3). For a user-given error bound $e$, suppose the setting I gets the compression result of \{bit rate= $B_{\textsc{I}}$, PSNR=$P_{\textsc{I}}$\}, and the setting II gets the compression result \{bit rate=$B_{\textsc{II}}$, PSNR=$P_{\textsc{II}}$\}. If $B_{\textsc{I}} > B_{\textsc{II}}$ and $P_{\textsc{I}} < P_{\textsc{II}}$, we can easily identify that II is better than I as II has lower bit rate (i.e., higher compression ratio) and higher PSNR (higher reconstructed data quality) meanwhile. However, if $B_{\textsc{I}} > B_{\textsc{II}}$ and $P_{\textsc{I}} > P_{\textsc{II}}$ or $B_{\textsc{I}} < B_{\textsc{\textsc{II}}}$ and $P_{\textsc{I}} < P_{\textsc{II}}$ (we call it sophisticated situation in the following text), additional analysis is needed to determine which solution is superior. 

We adopt an efficient online comparison method to determine the better choice for the sophisticated situation. Specifically, in this situation, QoZ uses another error bound (denoted as $e'$) which has a small offset to $e$ to perform a sampling-based trial compression for solution B, which can obtain another compression result (i.e., a pair of bit rate and quality metric (such as PSNR)): denoted as $B'_\textsc{II}$ and $P'_\textsc{II}$. Then, we can determine the better solution by checking the relationship between the solution I's result $(B_\textsc{I},P_\textsc{I})$ versus the line constructed by $(B_{\textsc{II}},P_{\textsc{II}})$ and $(B'_{\textsc{II}},P'_{\textsc{II}})$. If the $P_\textsc{I}$ is below the constructed line in space, QoZ asserts that the setting II is better, and vice versa. The second error bound $e'$ used for computing $B'_{\textsc{II}}$ and $P'_{\textsc{II}}$ is set to $1.2e$ if $P_{\textsc{I}} > P_{\textsc{II}}$ or $0.8e$ if $P_{\textsc{I}} < P_{\textsc{II}}$. Such a design can make $B_I$ lie in the range between $B_{\textsc{II}}$ and $B'_{\textsc{II}}$ in most of the cases based on our experience, obtaining a very accurate judgment accordingly.  


In Table \ref{tab:comparison}, we summarize all four situations for two comparative solutions I and II based on different compression results. Their compression results are denoted as \{$B_\textsc{I}$, $M_\textsc{I}$\} and \{$B_{\textsc{II}}$, $M_{\textsc{II}}$\}, respectively, where $M$ refers to the quality metric (such as PSNR, SSIM, AC). As shown in the table, we can directly identify the better solution for cases 1 and 2. For cases 3 and 4, an additional sampling-based trial compression run would be done for solution II with another error bound. Since all the trial compression runs are on top of sampled data, the computational overhead is very low, to be verified later. With this well-designed evaluation method, we can traverse all the candidate parameter sets and select the best one for practical compression.

\begin{table}[ht]
    \vspace{-1mm}
    \centering
    \caption{Comparison cases of two compression results $(B_I,M_I)$ and $(B_{II},M_{II})$ for solution I and II under the error bound $e$}  \vspace{-2mm}  
\resizebox{0.99\columnwidth}{!}{        
    \begin{tabular}{|c|c|c|}
    \hline
    Case\#& Case & Comparison  \\
    \hline
    1 & $B_\textsc{I}<=B_{\textsc{II}}$ and $M_{\textsc{I}}>=M_{\textsc{II}}$ & I is better\\
    \hline
    2 & $B_\textsc{I}>=B_{\textsc{II}}$ and $M_{\textsc{I}}<=M_{\textsc{II}}$ & II is better\\
    \hline
   \multirow{3}{*}{3} & \multirow{3}{*}{$B_{\textsc{I}}>B_{\textsc{II}}$ and $M_{\textsc{I}}>M_{\textsc{II}}$ } & compute $(B'_{\textsc{II}},M'_{\textsc{II}})$ with sol II and $0.8e$\\
    
    &  & draw a line with the 2 points from sol II\\
    &  &  check whether $(B_\textsc{I},M_\textsc{I})$ is above or below \\
   \hline
   \multirow{3}{*}{4} & \multirow{3}{*}{$B_{\textsc{I}}<B_{\textsc{II}}$ and $M_{\textsc{I}}<M_{\textsc{II}}$ } & compute $(B'_{\textsc{II}},M'_{\textsc{II}}) $with sol II and $1.2e$\\
    
    &  & draw a line with the 2 points from sol II  \\
    &  &  check whether $(B_\textsc{I},M_\textsc{I})$ is above or below \\
   \hline
    \end{tabular}}
    \label{tab:comparison}
\end{table}

%% file: tex/7_evaluation.tex
\section{Performance Evaluation}
\label{sec:evaluation}

In this section, we demonstrate the setup of our experiments and then present and analyze the experimental results. We evaluate our solution -- QoZ compared with four other state-of-the-art error-bounded lossy compressors, which have been verified as the leading related works by many prior studies \cite{Xin-bigdata18,sz3,zfp,liang2021mgard+,sz-auto,lu2018understanding}. We perform the evaluation based on multiple indicators comprehensively, including data distortion with various quality metrics, compression ratio, compression speed, visual quality, etc. The experiments are performed based on six well-known real-world scientific datasets across different domains, which are described as follows.

\subsection{Experimental Setup}

\subsubsection{Execution Environment}
The experiments of this paper are performed on the Argonne Bebop supercomputer which features over 2,000 nodes, and we used the bdwall nodes of it each having Intel Xeon E5-2695v4 CPU with 64 CPU cores and a total of 128GB of DRAM.

\subsubsection{Applications}

We evaluate the lossy compressors using six real-world scientific applications from different domains which are often used for the scientific data compression evaluation \cite{sdrb}:

\begin{itemize}
    \item CESM-ATM: CESM is a well-known climate simulation package, and we use its atmosphere model CESM-ATM \cite{sdrb} in our experiments. 
    \item RTM: Reverse time migration for seismic imaging \cite{geodriveFirstBreak2020}.
    \item NYX: A cosmological hydrodynamics simulation based on adaptive mesh \cite{nyx}.
    \item Hurricane: Simulation of Hurricane Isabel from the National Center for Atmospheric Research \cite{hurricane}.
    \item Scale-LETKF: Local Ensemble Transform Kalman Filter (LETKF) data assimilation package for the SCALE-RM weather model \cite{scale-letkf}.
    \item Miranda: Large-eddy simulation of multi-component flows by turbulent mixing via a radiation hydrodynamics code \cite{miranda}.
\end{itemize}

We detail the information about the datasets in Table~\ref{tab:dataset information}. 
\begin{table}[ht]
    \vspace{-1mm}
    \centering
    \caption{Information of the datasets in experiments}  \vspace{-2mm}  
\resizebox{0.99\columnwidth}{!}{  
    \begin{tabular}{|c|c|c|c|c|}
    \hline
    App.&\# fields& Dimensions & Total Size& Domain\\
    \hline
    RTM &3600&449$\times$449$\times$235&635GB&Seismic Wave\\
    \hline
    Miranda & 7 & 256$\times$384$\times$384& 1GB& Turbulence \\
    \hline
    CESM-ATM & 26 & 1800$\times$3600&1.5GB&Weather\\
    \hline
    Scale-LETKF & 13 & 98$\times$1200$\times$1200 & 6.4GB&Climate\\
    \hline
    NYX & 6 & 512$\times$512$\times$512 &3.1GB& Cosmology\\
    \hline
    Hurricane & 48$\times$13 & 100$\times$500$\times$500 &58GB& Weather\\
    \hline
    \end{tabular}}
    \label{tab:dataset information}
\end{table}

\subsubsection{Comparison of Lossy Compressors in Our Evaluation}

In our experiments, we compare QoZ with four other lossy compressors. The first one is SZ3 \cite{sz3}, which is a state-of-the-art error-bounded lossy compressor exhibiting very competitive compression quality in most cases. The second and third are SZ2.1 \cite{sz16,sz17,Xin-bigdata18} and ZFP0.5.5 \cite{zfp}, which have been widely used in the community. The last one is MGARD+ \cite{liang2021mgard+}, an improved version of MGARD \cite{MGARD}. We use MGARD+ instead of MGARD in that MGARD+ can achieve both significantly higher performance and compression quality over MGARD according to \cite{liang2021mgard+}.

\subsubsection{Experimental Configurations}

In our implementation, for 2D input data, the sample block size and anchor point stride are set to both 64, and for 3D input data, the sample block size and anchor point stride are set to 16 and 32. For interpolation/parameter selection, we sample 1\% of the input for 2D data or 0.5\% of the input for 3D data. For other compressors including SZ2.1, ZFP0.5.5, SZ3, and MGARD+, we use their default configurations for fairness of comparison.


\subsubsection{Evaluation Metrics}

We perform the evaluation based on five critical metrics: 
\begin{itemize}
    \item Compression ratio (CR) under the same error bound: We have described CR and error bound in Section \ref{sec:problemform}. We adopt value-range-based error bound (denoted by $\epsilon$), as it has been widely used in the lossy compression community \cite{Xin-bigdata18,sz3,liang2021mgard+,sz-auto}. It is essentially equivalent to the absolute error bound (denoted $e$), as $e$ = $\epsilon \cdot value\_range$. 
    \item \textit{Rate-PSNR plots}: A critical rate-distortion evaluation used in lossy compression (see Section \ref{sec:problemform} for details).
    \item \textit{Rate-SSIM plots}: Another rate distortion evaluation in which SSIM is commonly used to assess visual quality \cite{ssim} and reconstructed data quality in scientific analysis \cite{Baker-Climate17,dssim}. 
    \item \textit{rate-AC}: Autocorrelation of compression errors is a concerned metric for many applications, as discussed in Section \ref{sec:problemform}.
    \item Visualization with the same CR: We compare the visual quality of the reconstructed data based on the same CR.
    \item Compression/decompression speed: We check the overall compression/decompression speed of our QoZ framework to show the low computational overhead in our solution.  
    \item Parallel I/O performance: We perform a parallel data transfer evaluation on a supercomputer.  
\end{itemize}

\subsection{Evaluation Results and Analysis}

\subsubsection{Verification of compression errors versus error bound}

\begin{figure}[ht] 
\centering
\hspace{-12mm}
\subfigure[{CESM-ATM (CLDHGH field)}]
{
\raisebox{-1cm}{\includegraphics[scale=0.3]{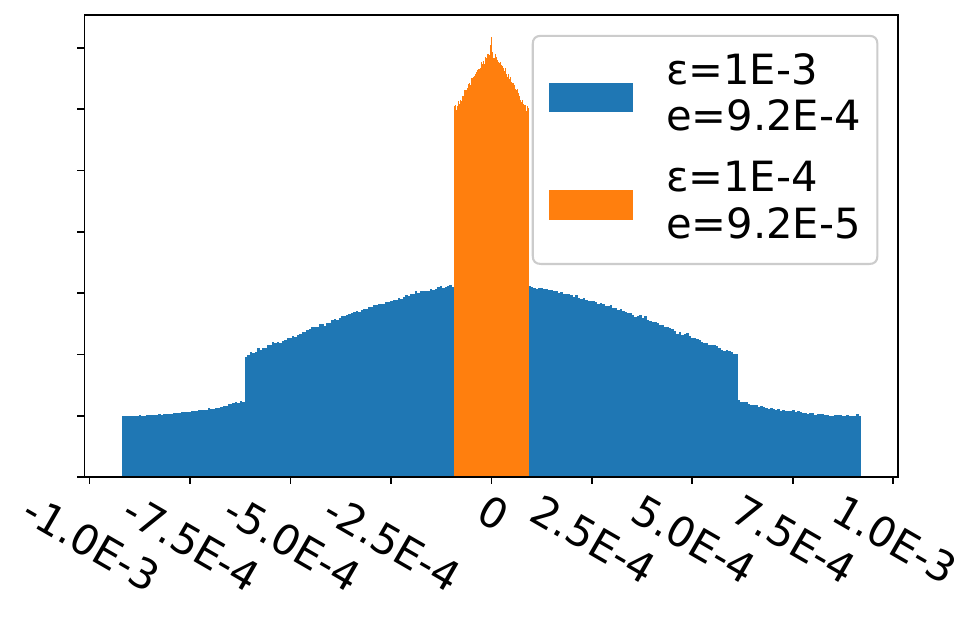}}
}%
\hspace{-7mm}
\subfigure[{NYX (Baryon density field)}]
{
\raisebox{-1cm}{\includegraphics[scale=0.3]{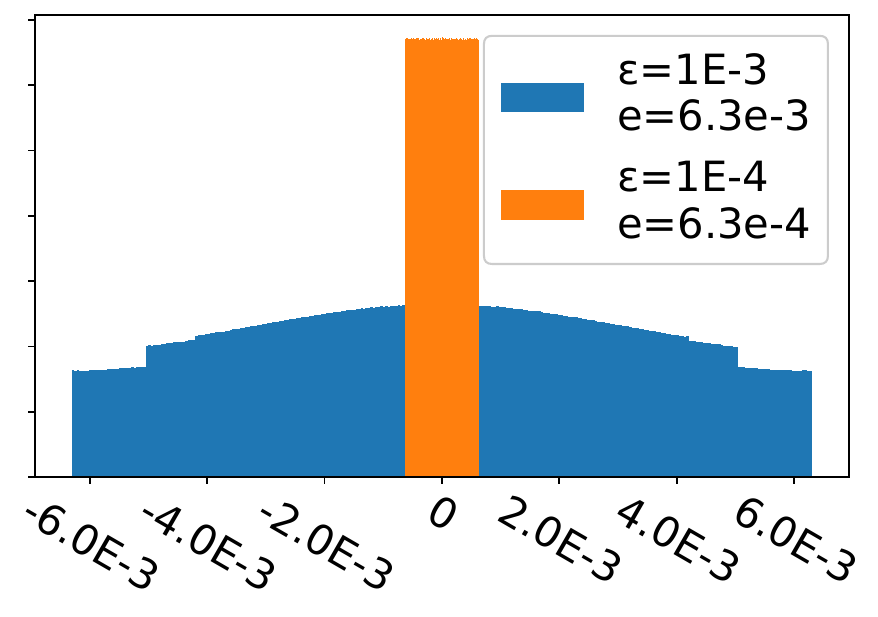}}%
}
\hspace{-12mm}
\vspace{-1mm}

\caption{Distribution of Compression Errors}
\label{fig:err-verification}
\end{figure}

First of all, we verified the compression errors for QoZ based on each experimental dataset under different error bounds. We confirm that QoZ always strictly respects the user-specified error bounds, as exemplified in Figure \ref{fig:err-verification}: the distribution of compression errors for QoZ based on two application datasets with different error bounds. We can clearly observe that the compression errors are always confined within the error bound ($e$) for all data points.

\subsubsection{Compression ratio under the same error bounds}

We compare the compression ratios of all lossy compressors under the same certain error bounds, in which QoZ's tuning mode is set to `maximizing compression ratio' (i.e., the compression ratio is most concerned). Table \ref{tab:compress ratio comparison} shows the compression ratios of the 5 lossy compressors on the 6 datasets under 3 error bounds (1e-2, 1e-3, 1e-4). It can be concluded that QoZ has the leading compression ratios in most of the cases, though the improvement is not significant in some cases. 
In particular, under the error bound of 1e-2, QoZ's compression ratio is 70\% higher than that of SZ3 on the Miranda dataset; and 20\% higher on the RTM dataset. 

\begin{table}[ht]
\centering
\footnotesize
  \caption {Compression Ratio Comparison Based on the Same Error Bound} 
  \vspace{-2mm}
  \label{tab:compress ratio comparison} 
  \begin{adjustbox}{width=\columnwidth}
  \begin{tabular}{|c|c|c|c|c|c|c|c|}
  \hline
    \multirow{2}{*}{Dataset} & \multirow{2}{*}{$\epsilon$}  & SZ & SZ & \multirow{2}{*}{ZFP} & \multirow{2}{*}{MGARD+}  & QoZ & OurSol \\ 
     &  & 2.1 & 3 & &  & (OurSol) & Improve \% \\ \hline
\multirow{3}{*}{RTM}  &  1E-2  & 283.3  & 2041.6 & 110.9 & 234.2 & \textbf{2461.2} & 20.6\%  \\ \cline{2 - 8}
   &  1E-3  & 106.8 &  417.0 & 59.2 & 78.5 & \textbf{447.8} & 7.4\% \\ \cline{2 - 8}
   &  1E-4  & 54.4  & 118.1  & 35.0 & 38.3 & \textbf{122.3} & 3.6\%\\ \cline{1 - 8}
\multirow{3}{*}{Miranda}  &  1E-2  &  126.3  & 574.6 & 46.6 &52.1 & \textbf{987.2} & 71.8\%  \\ \cline{2 - 8}
  &  1E-3  & 59.5 & 168.0  & 25.6& 26.2& \textbf{177.1} & 5.5\%  \\ \cline{2 - 8}
  &  1E-4  & 29.6  &  47.3 & 14.5 & 12.9 & \textbf{48.2} & 1.9\% \\ \cline{1 - 8}
\multirow{3}{*}{CESM-ATM}  &  1E-2  &  151.5  & \textbf{380.9} &7.7 & 49.3 & 373.8 &  -1.9\% \\ \cline{2 - 8}
  &  1E-3  &38.3 & 57.6 &5.4 & 20.8& \textbf{60.2} & 4.4\% \\ \cline{2 - 8}
  &  1E-4  & 14.5  & 16.6 & 4.0 & 9.5 & \textbf{16.9} & 2.2\%  \\ \cline{1 - 8}
\multirow{3}{*}{SCALE-LKF}  &  1E-2  & 84.0   & \textbf{167.3} & 14.5 &  53.8 & 163.4 & -2.3\% \\ \cline{2 - 8}
  &  1E-3  & 26.5 & 40.4 & 7.8& 20.3 & \textbf{41.8} & 3.4\% \\ \cline{2 - 8}
  &  1E-4  & 13.9  & \textbf{14.1} & 4.6 & 10.4 & 13.4 & -5.3\%  \\ \cline{1 - 8}
\multirow{3}{*}{NYX}  &  1E-2  & 44.1 & 61.3& 12.0 & 24.7 &\textbf{62.0}  & 1.2\% \\ \cline{2 - 8}
  &  1E-3  & 17.1  & 21.5  & 6.0 & 11.2 & \textbf{21.7} & 1.3\% \\ \cline{2 - 8}
  &  1E-4  &  7.7 & 9.1 & 3.7 &5.5 & \textbf{9.2}  & 0.7\%  \\ \cline{1 - 8}
\multirow{3}{*}{Hurricane}  &  1E-2  &  49.8  & 69.0 & 11.3& 28.4 &\textbf{70.3} & 1.8\% \\ \cline{2 - 8}
  &  1E-3  &  17.5 & 21.8 & 6.7 & 12.7 &  \textbf{22.2} & 1.7\%   \\ \cline{2 - 8}
  &  1E-4  & 9.8  & \textbf{10.5}  & 4.3 & 7.4 & 9.3 & -12.6\% \\ \cline{1 - 8}
\end{tabular}
\end{adjustbox}
\end{table}

\subsubsection{Rate distortion evaluation}

In the following text, we present the overall rate-distortion results of QoZ versus other state-of-the-art error-bounded lossy compressors, in regard to different quality metrics. 

\begin{figure}[ht] 
\centering
\hspace{-10mm}
\subfigure[{RTM}]
{
\raisebox{-1cm}{\includegraphics[scale=0.39]{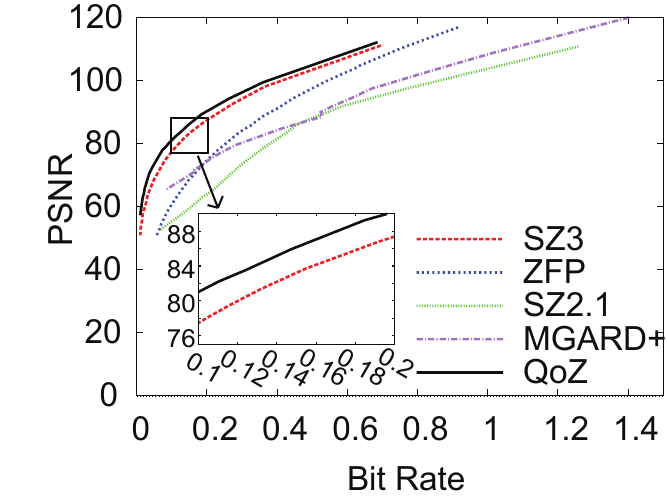}}
}
\hspace{-6mm}
\subfigure[{Miranda}]
{
\raisebox{-1cm}{\includegraphics[scale=0.39]{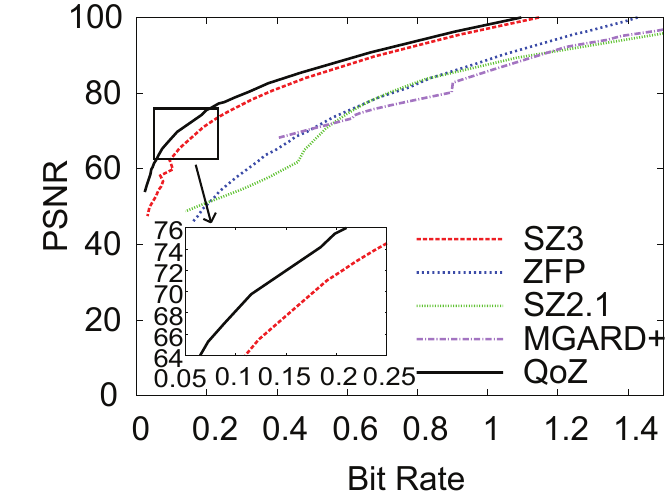}}
}
\hspace{-10mm}
\vspace{-1mm}

\hspace{-10mm}
\subfigure[{CESM-ATM}]
{
\raisebox{-1cm}{\includegraphics[scale=0.39]{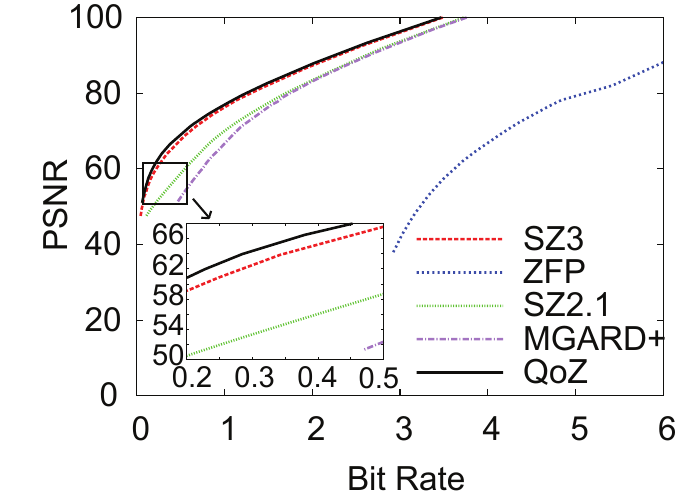}}
}
\hspace{-6mm}
\subfigure[{SCALE-LETKF}]
{
\raisebox{-1cm}{\includegraphics[scale=0.39]{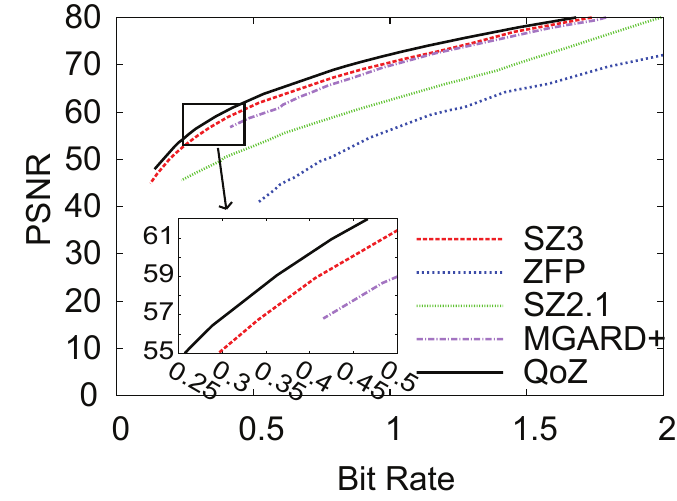}}%
}
\hspace{-10mm}
\vspace{-1mm}

\hspace{-10mm}
\subfigure[{NYX}]
{
\raisebox{-1cm}{\includegraphics[scale=0.39]{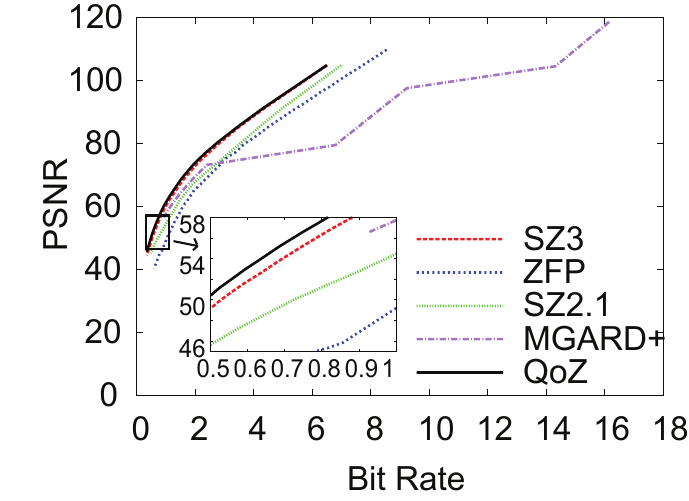}}%
}
\hspace{-6mm}
\subfigure[{Hurricane-Isabel}]
{
\raisebox{-1cm}{\includegraphics[scale=0.39]{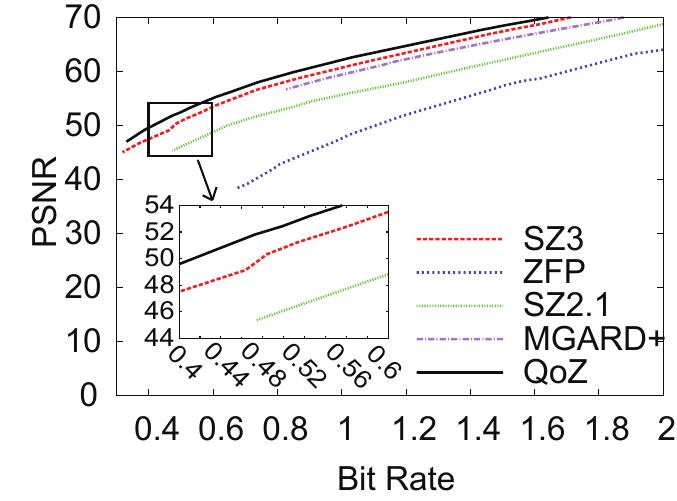}}%
}
\hspace{-10mm}
\vspace{-1mm}

\caption{Rate Distortion Evaluation (PSNR)}
\label{fig:rate-psnr}
\end{figure}

Figure \ref{fig:rate-psnr} shows the rate-PSNR evaluation results of each lossy compressor on all datasets in which QoZ applies the tuning mode of `rate-PSNR preferred'. In the plots, the x-axis is bit rate and the y-axis is PSNR. We can observe that QoZ achieves the best rate-distortion curve on all the datasets. For example, QoZ achieves $\sim$150\%/$\sim$70\% improvement in compression ratio on the Miranda dataset than the second-best (SZ3) when PSNR is around 55/65, and $\sim$80\% improvement in compression ratio on the RTM dataset when PSNR is $\sim$60. QoZ also provides $\sim$20\% improvement in compression ratio on SCALE-LETKF dataset than SZ3 at a PSNR of 55. 
The reason QoZ achieves better rate-PSNR over SZ3 is mainly 3-fold: First, the level-wise error bounds in SZ3 are fixed for the entire dataset, but QoZ uses an auto-tuning algorithm to dynamically set appropriate error bounds for each level, according to preferred quality metric -- PSNR. Second, QoZ fine-tunes different interpolators on different prediction levels to improve the prediction accuracy. Last, the anchor points in QoZ eliminate the long-range prediction in SZ3, which significantly increases the prediction accuracy on inputs having various data patterns in different regions (especially for Miranda and RTM datasets). 

\begin{figure}[ht] 
\centering
\hspace{-10mm}
\subfigure[{RTM}]
{
\raisebox{-1cm}{\includegraphics[scale=0.39]{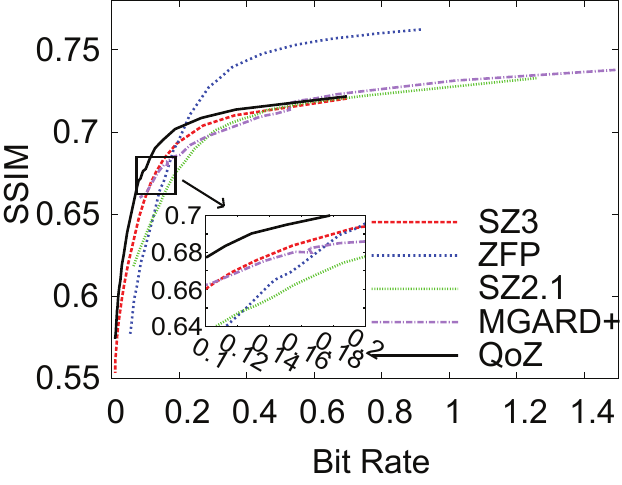}}%
}
\hspace{-5mm}
\subfigure[{Miranda}]
{
\raisebox{-1cm}{\includegraphics[scale=0.39]{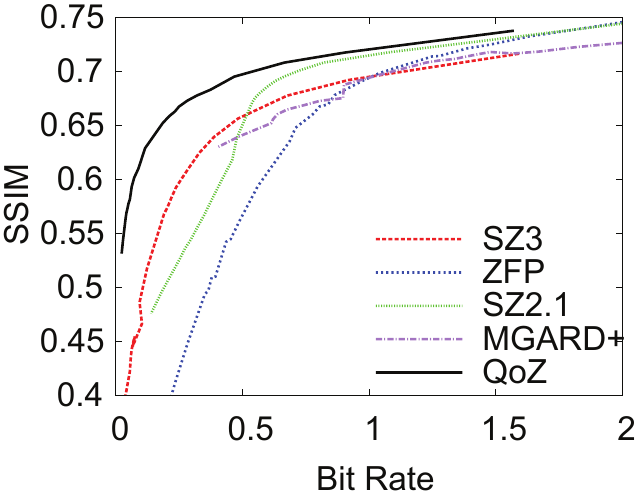}}%
}
\hspace{-10mm}
\vspace{-1mm}

\hspace{-10mm}
\subfigure[{CESM-ATM}]
{
\raisebox{-1cm}{\includegraphics[scale=0.39]{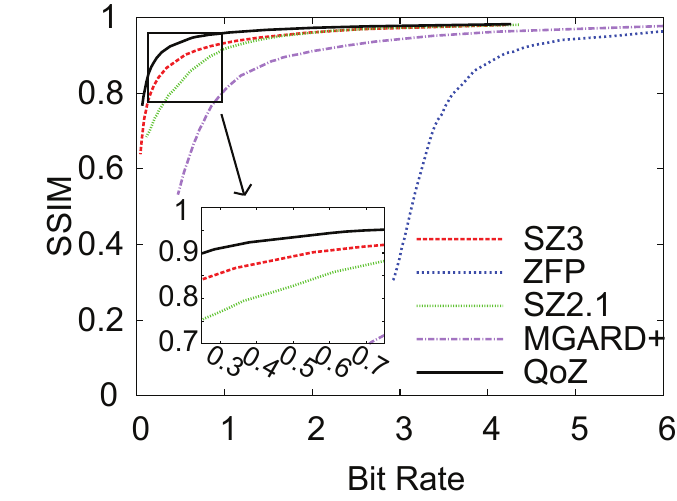}}%
}
\hspace{-5mm}
\subfigure[{SCALE-LETKF}]
{
\raisebox{-1cm}{\includegraphics[scale=0.39]{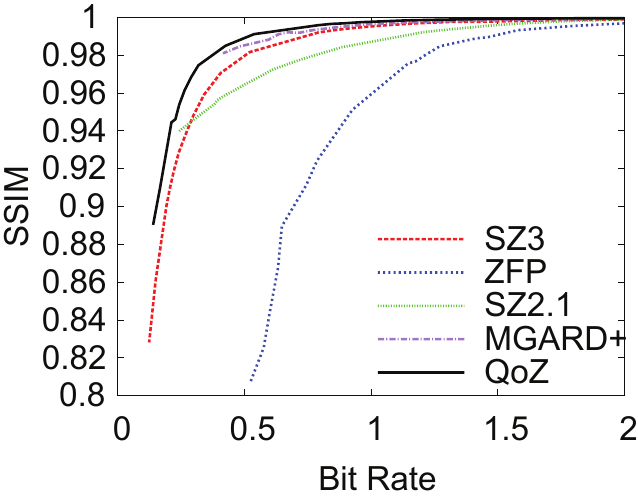}}%
}
\hspace{-10mm}
\vspace{-1mm}

\hspace{-10mm}
\subfigure[{NYX}]
{
\raisebox{-1cm}{\includegraphics[scale=0.39]{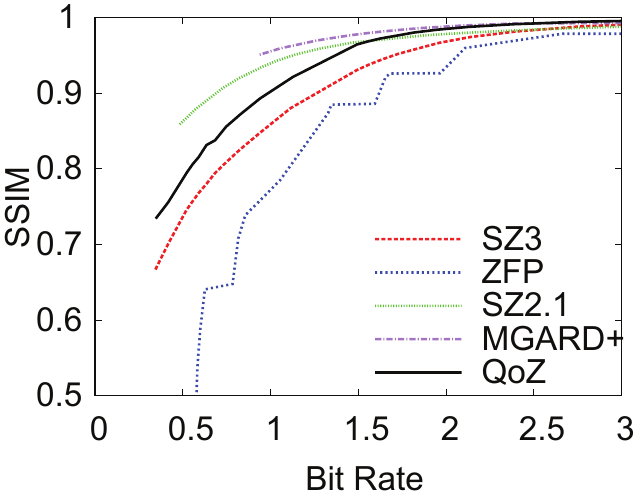}}%
}
\hspace{-6mm}
\subfigure[{Hurricane-Isabel}]
{
\raisebox{-1cm}{\includegraphics[scale=0.39]{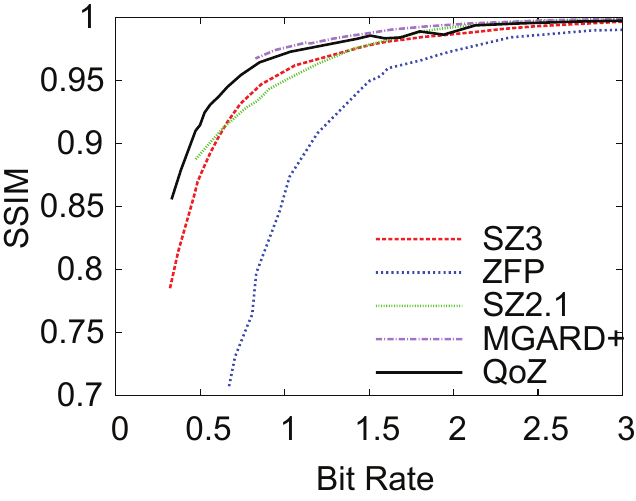}}%
}
\hspace{-10mm}
\vspace{-1mm}

\caption{Rate Distortion Evaluation (SSIM)}
\label{fig:rate-ssim}
\end{figure}

Figure \ref{fig:rate-ssim} shows the rate-SSIM plots of different lossy compressors: the more top-left the curve approaches, the better the result. We observe that our AoZ has the best (or near best) rate-SSIM in most of cases for different datasets. In absolute terms, when SSIM reaches 0.9 on the CESM-ATM dataset, QoZ has $\sim$120\% compression ratio improvement over the second-best solution (SZ3). On the Miranda dataset, when SSIM is around 0.6/0.65, QoZ's compression ratio is about 270\%/150\% higher than that of the second best (SZ3). For the Hurricane-Isabel dataset, QoZ obtains 25\% gains in compression ratio over the second best (SZ2) when SSIM is around 0.9. Another interesting observation is that QoZ is not always the best at all bit-rates in the rate-SSIM plots, which implies that PSNR and SSIM are two distinct important metrics, which cannot be substituted with each other in practice. 

\subsubsection{Autocorrelations}

\begin{figure}[ht] 
\centering



\hspace{-10mm}
\subfigure[{RTM}]
{
\raisebox{-1cm}{\includegraphics[scale=0.39]{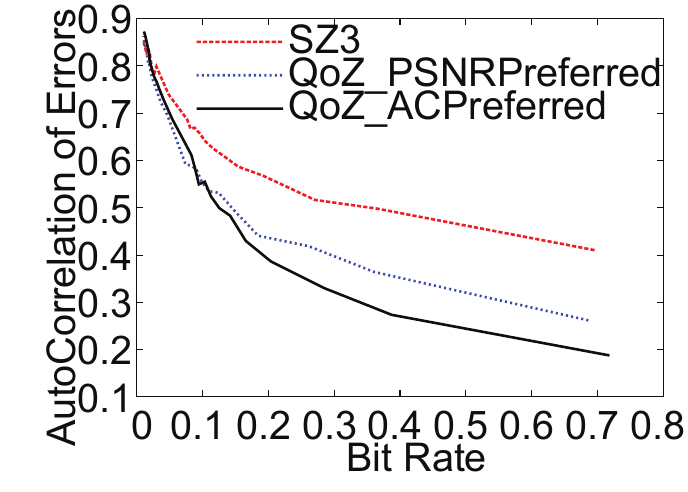}}%
}
\hspace{-5mm}
\subfigure[{Miranda}]
{
\raisebox{-1cm}{\includegraphics[scale=0.39]{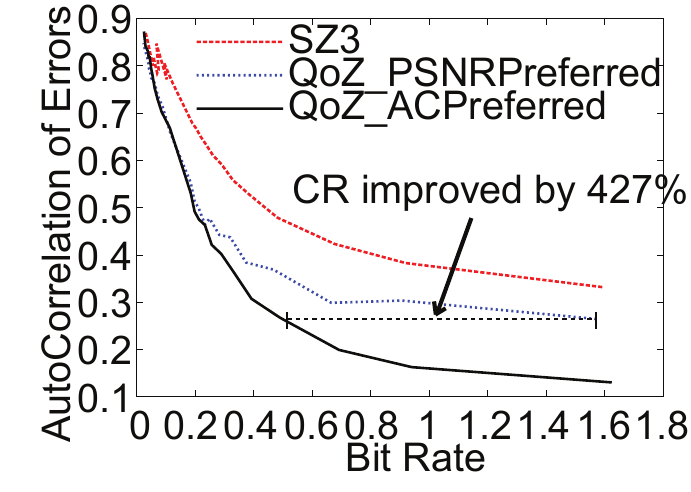}}%
}
\hspace{-10mm}
\vspace{-1mm}

\hspace{-10mm}
\subfigure[{CESM-ATM}]
{
\raisebox{-1cm}{\includegraphics[scale=0.39]{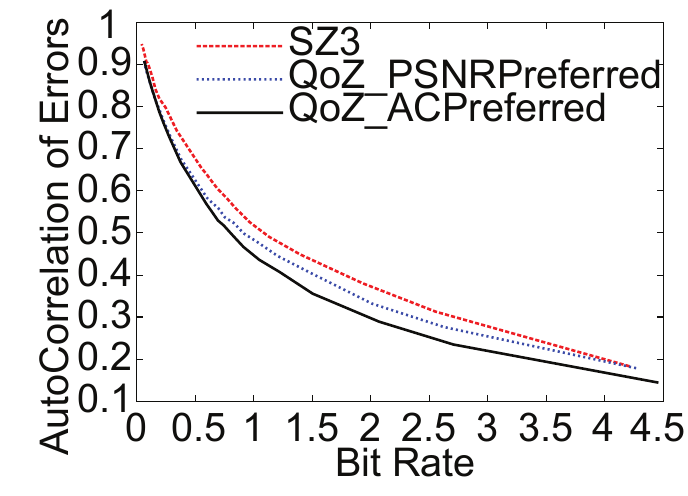}}%
}
\hspace{-5mm}
\subfigure[{SCALE-LETKF}]
{
\raisebox{-1cm}{\includegraphics[scale=0.39]{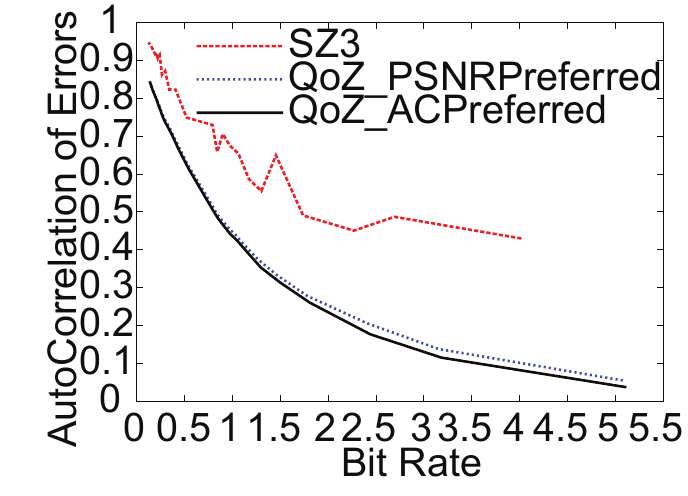}}%
}
\hspace{-10mm}
\vspace{-1mm}

\hspace{-10mm}
\subfigure[{NYX}]
{
\raisebox{-1cm}{\includegraphics[scale=0.39]{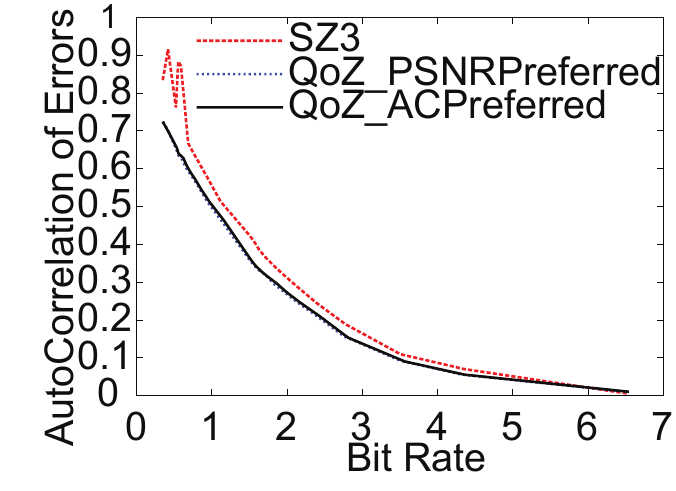}}%
}
\hspace{-6mm}
\subfigure[{Hurricane-Isabel}]
{
\raisebox{-1cm}{\includegraphics[scale=0.39]{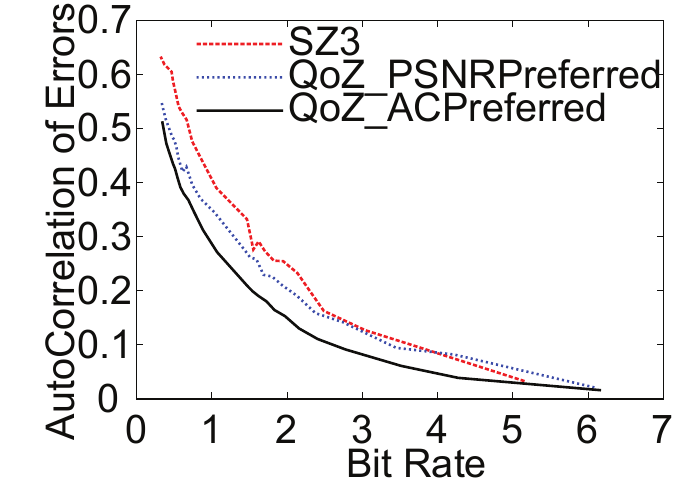}}%
}
\hspace{-10mm}
\vspace{-1mm}
\caption{Rate Distortion Evaluation (AutoCorrelation of Errors)}
\label{fig:rate-ac}
\end{figure}

Figure \ref{fig:rate-ac} presents rate-autocorrelation (Lag 1 autocorrelation of errors) for SZ3, QoZ (PSNR preferred mode), and QoZ (autocorrelation preferred mode) on all the 6 datasets. As lower autocorrelation means compression errors are more randomly distributed and therefore better, we can observe that QoZ always has better autocorrelations than SZ3 under the same bit rate no matter which mode is used, especially on RTM, Miranda, and SCALE-LETKF datasets. Moreover, when using the autocorrelation preferred mode, the autocorrelation of QoZ is further improved than using the PSNR preferred mode, which confirms the importance and effectiveness of our quality-metric-oriented design in QoZ. In particular, on the Miranda dataset, QoZ with the AC preferred mode can achieve up to 427\% improvement in compression ratio over the compression result with the PSNR preferred mode.

\begin{figure}[ht]
  \centering
  \raisebox{-1mm}{\includegraphics[scale=0.45]{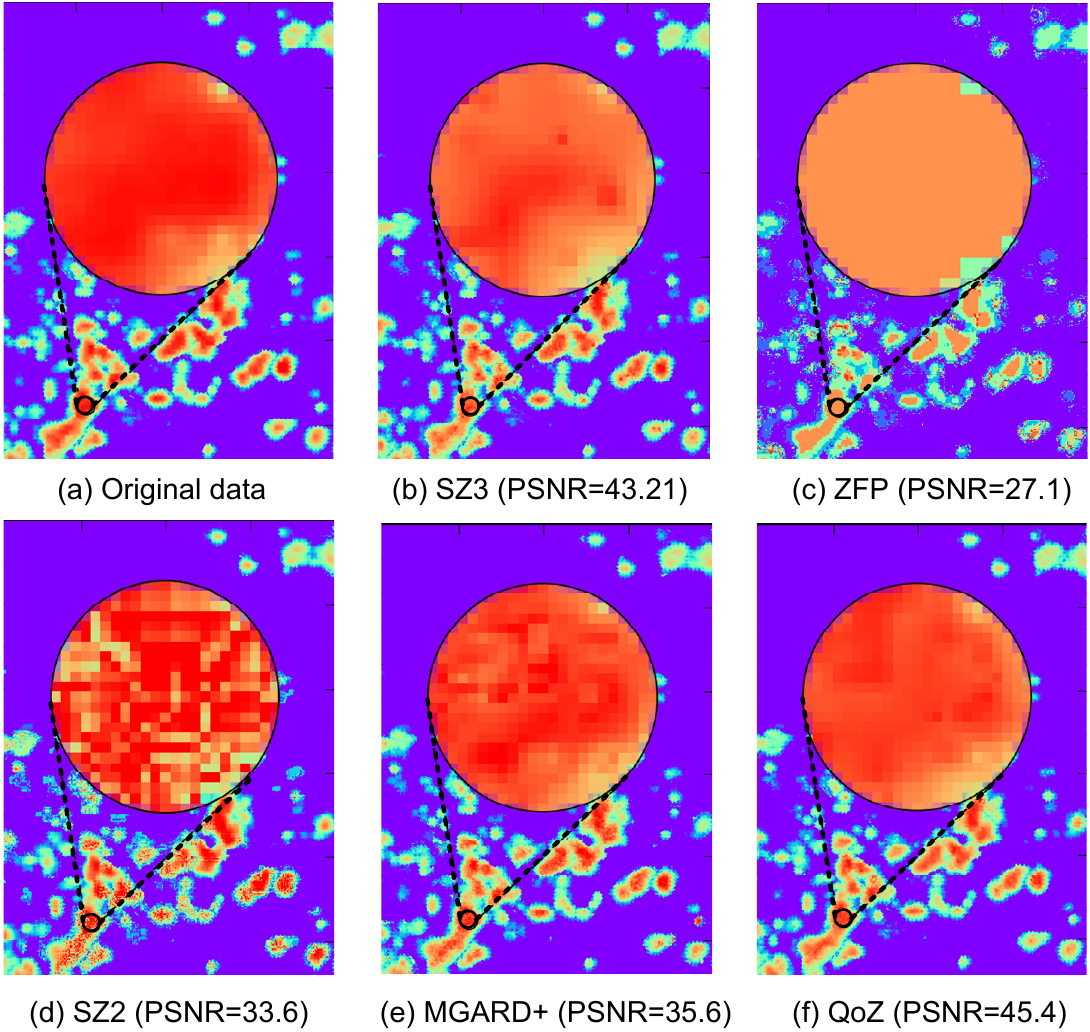}}%
  \vspace{-2mm}
  \caption{Visualization of Reconstructed Data with CR=65  (Scale-LETKF)}
  \label{fig:Scale-vis}
\end{figure}

\subsubsection{Decompression data visualizations}

We demonstrate the high visual quality of QoZ in figure \ref{fig:Scale-vis}, as compared with other compressors under the same compression ratio (65) based on the SCALE-LETKF dataset. It is observed that QoZ has the highest PSNR visual quality (with a tiny data distortion though), which is mainly due to the level-adapted prediction design and parameter fine-tuning at different levels. 
\begin{figure}[ht] 
\centering
\hspace{-10mm}
\subfigure[{CESM-ATM}]
{
\raisebox{-1cm}{\includegraphics[scale=0.4]{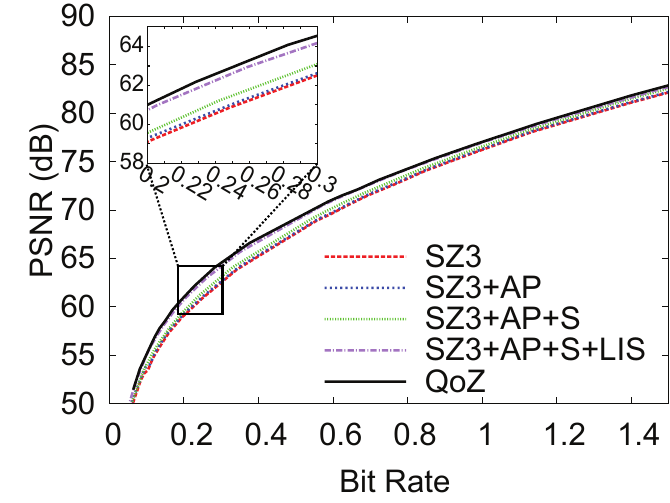}}%
}
\hspace{-6mm}
\subfigure[{Miranda}]
{
\raisebox{-1cm}{\includegraphics[scale=0.4]{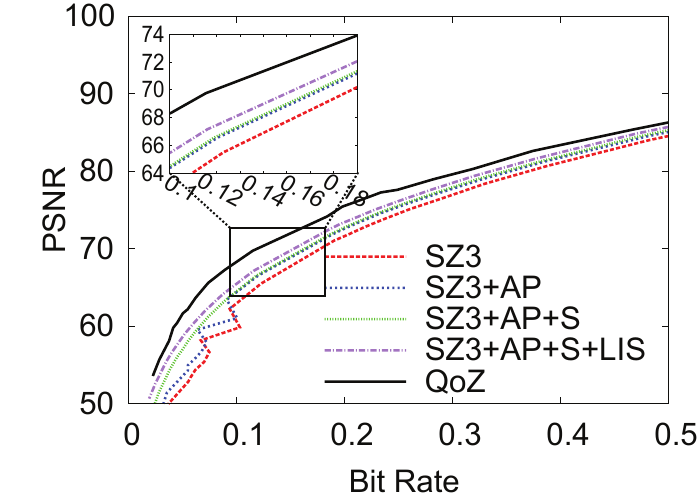}}%
}
\hspace{-10mm}
\vspace{-1mm}
\caption{Ablation Study for The Compression Quality (rate-distortion (PSNR)) Contributed by Different Components}
\label{fig:component-rate-psnr}
\end{figure}

\begin{figure}[ht] 
\centering
\hspace{-10mm}
\subfigure[{CESM-ATM}]
{
\raisebox{-1cm}{\includegraphics[scale=0.4]{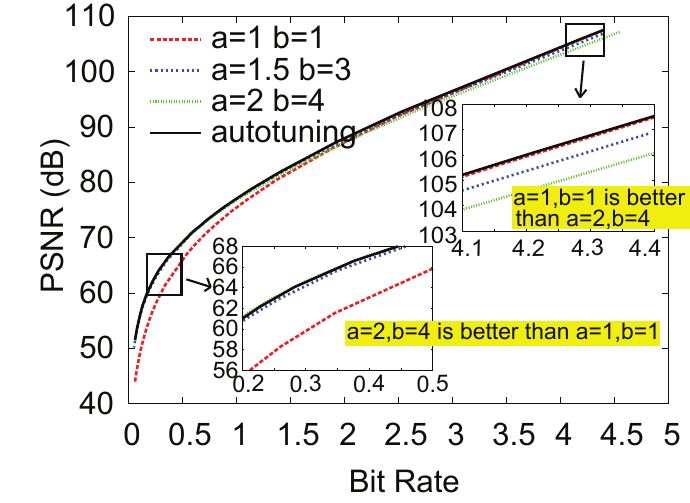}}%
}
\hspace{-6mm}
\subfigure[{NYX}]
{
\raisebox{-1cm}{\includegraphics[scale=0.4]{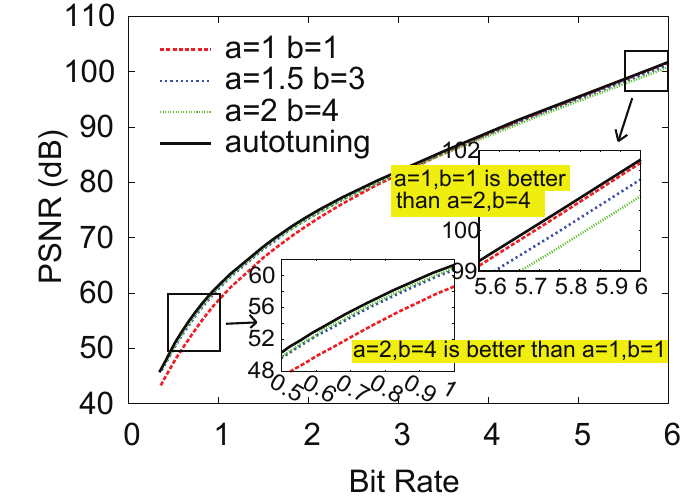}}%
}
\hspace{-10mm}
\vspace{-1mm}

\caption{Analysis of Impact of Parameter Tuning on Compression Quality}
\label{fig:ab-rate-psnr}
\end{figure}

\subsubsection{Ablation Study}

As shown before, the design of QoZ is based on several different new components, including anchor points, level-wise interpolator/error bound, metric-driven auto-tuning, etc. To verify and illustrate how each single design component in QoZ takes effect and contribute to the compression with QoZ, we perform an ablation study to understand the various contribution of different components in our design to the overall compression quality. Specifically, we construct different compression methods/versions by adding our designed components one by one and observing the change in overall compression ratios. The decomposed design components in this study include the design of anchor point (denoted as \textit{AP}), our designed uniform sampling method (denoted as \textit{S}), optimized level-based interpolation selection (denoted as \textit{LIS}) and parameter autotuning (PA). As such, there are five compression methods in total: SZ3 is the original baseline version \cite{sz3}; SZ3+AP is applying AP over SZ3; SZ3+AP+S is further applying our improved uniform sampling on SZ3+AP to select the best-fit interpolation method for an overall dataset; SZ3+AP+S+LIS is adopting fine-grained selection of interpolation method for different levels on top of SZ3+AP+S; QoZ is our final solution with all techniques including AP, S, LIS and PA. Due to the space limit, we demonstrate the evaluation results using only two typical application datasets (CESM-ATM and Miranda). As shown in Figure \ref{fig:component-rate-psnr}, by adding our designed components to SZ3, the rate-distortion keeps increasing. Our final solution -- QoZ exhibits the best rate-distortion (see the black curve) in class, which verifies the importance and effectiveness of each of our designed optimization components.

Then, we analyze the lossy compression qualities of QoZ with different parameter tuning methods to verify the effectiveness of our parameter auto-tuning algorithm. As mentioned previously, there are two critical parameters (i.e., $\alpha$ and $\beta$) to tune for controlling the error propagation in our multi-level interpolation-based predictor (see Formula \ref{eq:ebl}). Figure \ref{fig:ab-rate-psnr} shows the different rate-distortion (w.r.t. PSNR) generated by QoZ using different parameter settings on 2 datasets (CESM-ATM and NYX). These results verify that the best parameter settings often alter at different bit rates (i.e., compression ratios), which indicates a strong motivation for our parameter auto-tuning design. As verified in the figure, our designed parameter auto-tuning method can always lead to the best rate distortion at different bit rates. 



\subsubsection{Compression/decompression Speeds}

We present both compression and decompression speed in Table \ref{tab:speed}, under the error bound of 1e-3, in which QoZ applies the PSNR preferred mode. QoZ keeps comparable performances with SZ3 and MGARD+. This verifies the low computational overhead of our quality metric oriented design, which is mainly attributed to the efficient uniform sampling method (Section \ref{sec:sampling}) and the narrowed parameter candidates (\ref{sec:constructing}). Although SZ2 and ZFP has higher speed than QoZ does, they suffer significantly worse rate distortions (as shown in Figure \ref{fig:rate-psnr} and \ref{fig:rate-ssim}).

\begin{table}[ht]
\centering
\footnotesize
  \caption {Compression/Decompression Speeds (MB/s) with $\epsilon$=1e-3 } 
  \vspace{-2mm}
  \label{tab:speed} 
  \begin{adjustbox}{width=\columnwidth}
  \begin{tabular}{|c|c|c|c|c|c|c|}
  \hline
      \multirow{2}{*}{Type} & \multirow{2}{*}{Dataset} & SZ & SZ & \multirow{2}{*}{ZFP} & \multirow{2}{*}{MGARD+}  & QoZ  \\ & & 2.1 & 3 &   &  & (OurSol)  \\ \hline
    \multirow{6}{*}{\rotatebox[origin=c]{90}{Compression}} & RTM & 207 & 147  & 556 & 142 & 129    \\ \cline{2-7}
    & Miranda & 201  & 134 & 239&149 &124   \\ \cline{2-7}
    & CESM-ATM & 181 &127 &194 & 125 &133   \\ \cline{2-7}
    & SCALE-LETKF &158 &135 &131 & 143& 131  \\ \cline{2-7}
    & NYX & 181& 98& 149& 131 &97  \\ \cline{2-7}
    & Hurricane & 159 & 127& 137& 152 &119  \\ \cline{1-7}
    \multirow{6}{*}{\rotatebox[origin=c]{90}{Decompression}} & RTM &452& 410& 996& 210& 388    \\ \cline{2-7}
    & Miranda &404 &374 &659 &212 &350   \\ \cline{2-7}
    & CESM-ATM &344 &381& 246& 203& 375  \\ \cline{2-7}
    & SCALE-LETKF &305 & 359& 362& 202 & 342 \\ \cline{2-7}
    & NYX & 281&172& 281& 148& 169 \\ \cline{2-7}
    & Hurricane & 266& 279& 321& 196& 278   \\ \cline{1-7}
\end{tabular}
\end{adjustbox}

\vspace{3mm}
\end{table}

\begin{figure}[ht]
\centering
\raisebox{-2cm}{\includegraphics[scale=0.5]{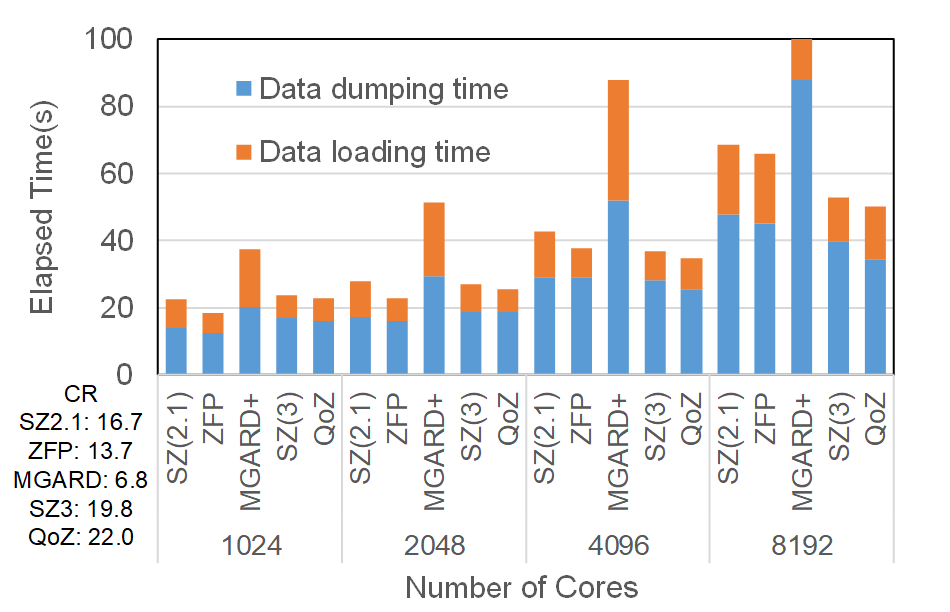}}

\vspace{3mm}
\caption{Parallel Performance Evaluation of Hurricane-Isabel Simulation (Compression ratios shown on bottom-left)}
\label{fig:parallel-hurricane}
\end{figure}

\subsubsection{Parallel Data Dumping/Loading Performance}

We evaluate the data dumping and loading performance of the Hurricane-Isabel simulation with different lossy compressors using 1K$\sim$8K cores (each core possessing a fixed amount of data data 1.3GB). Figure \ref{fig:parallel-hurricane} shows that QoZ provides the best overall performance especially when the total data size reaches several terabytes (over 5TB in total), which is attributed to the best compression ratio under QoZ and limited I/O bandwidth of the Bebop supercomputer. 

%% file: tex/8_conclusion.tex
\section{Conclusion and Future Work}
\label{sec:conclusion}

We propose QoZ, a quality metric oriented error-bounded compressor. The critical feature of our design is being able to flexibly adjust rate distortion based on various quality metrics, also respecting error bound constraints. We evaluate our solution and other state-of-the-art compressors with 6 real-world datasets on up to 8K cores. The key observations are summarized as follows. 
\begin{itemize}
    \item QoZ locates and analyzes the limitations of the interpolation-based data predictor in SZ3, then presents a new design to overcome those limitations and adapt to the quality metric driven lossy compression.
    \item Compared with the second best compressor, QoZ gains up to 70\% improvement in compression ratio under the same error bound, up to 150\% compression ratio improvement under the same PSNR, or up to 270\% compression ratio improvement under the same SSIM. QoZ also improves autocorrelation over SZ3 in all cases.
    \item QoZ has the best visual quality on reconstructed data with the same compression ratio among all tested compressors. 
    \item QoZ has comparable sequential compression/decompression speeds with the other state-of-the-art compressors and achieves the leading data dumping/loading performance on large-scale parallel running.
\end{itemize}
Currently, QoZ has a few limitations. First, its running speed is still lower than SZ2.1 and ZFP. Second, the interpolation types included in QoZ could be improved (currently including only linear and cubic spline interpolations). Third, dynamically adjusting hyper-parameters in QoZ leads to a high adaptability for QoZ, while the hyper-parameter selection policy stems from empirical studies. In future work, we will refine QoZ in several ways such as developing better predictors, designing better predictor selection policy and hyper-parameter auto-tuning methods, and improving QoZ's running speed.